\documentclass[journal]{IEEEtran}
\usepackage{amsmath,amssymb,graphicx}
\usepackage{color}
\usepackage{cite}
\usepackage{balance}

\usepackage[]{algorithm2e}
\usepackage{wasysym}
\usepackage{bm}
\usepackage{balance}
\usepackage{enumitem} %
\usepackage{pifont}%
\newcommand{\batsu}{\ding{53}}%

\def\T{{\mathsf T}}

\def\RR{\mathbb R}

\def\F{{\mathrm F}}
\def\T{{\mathsf T}}

\DeclareMathOperator*{\argmin}{argmin}
\DeclareMathOperator*{\argmax}{argmax}
\DeclareMathOperator{\real}{Re}
\renewcommand{\j}{{\mathrm{j}}}

\def\vec#1{\ensuremath{\bm{{#1}}}}

\begin{document}
\title{Phasebook and Friends:\\Leveraging discrete representations\\for source separation}
\author{Jonathan Le Roux,~\IEEEmembership{Senior Member,~IEEE,} 
Gordon Wichern,~\IEEEmembership{Member,~IEEE,}
Shinji Watanabe,~\IEEEmembership{Senior Member,~IEEE,}
Andy Sarroff,~\IEEEmembership{Member,~IEEE,}
John R. Hershey,~\IEEEmembership{Senior Member,~IEEE}%
\thanks{J. Le Roux and G. Wichern are with Mitsubishi Electric Research Laboratories (MERL), Cambridge, MA, USA (e-mail: \{leroux,wichern\}@merl.com). S. Watanabe is with Johns Hopkins University, Baltimore, MD, USA (e-mail: shinjiw@ieee.org). A. Sarroff is with iZotope, Cambridge, MA, USA (e-mail: asarroff@izotope.com). J. R. Hershey is with Google, Cambridge, MA, USA (e-mail: johnhershey@google.com).}%
\thanks{All authors contributed to this work while they were at MERL.}
}%

\markboth{}%
{Le Roux \MakeLowercase{\textit{et al.}}: Phasebook and Friends}

\maketitle

\begin{abstract}
Deep learning based speech enhancement and source separation systems have recently reached unprecedented levels of quality, to the point that performance is reaching a new ceiling. Most systems rely on estimating the magnitude of a target source by estimating a real-valued mask to be applied to a time-frequency representation of the mixture signal. A limiting factor in such approaches is a lack of phase estimation: the phase of the mixture is most often used when reconstructing the estimated time-domain signal. %
Here, we propose ``magbook'', ``phasebook'', and ``combook'', three new types of layers based on discrete representations that can be used to estimate complex time-frequency masks. Magbook layers extend classical sigmoidal units and a recently introduced convex softmax activation for mask-based magnitude estimation. Phasebook layers use a similar structure to give an estimate of the phase mask without suffering from phase wrapping issues. Combook layers are an alternative to the magbook-phasebook combination that directly estimate complex masks.
We present various training and inference schemes involving these representations, and explain in particular how to include them in an end-to-end learning framework.
We also present an oracle study to assess upper bounds on performance for various types of masks using discrete phase representations. We evaluate the proposed methods on the wsj0-2mix dataset, a well-studied corpus for single-channel speaker-independent speaker separation, matching the performance of state-of-the-art mask-based approaches without requiring additional phase reconstruction steps.
\end{abstract}


\IEEEpeerreviewmaketitle

\section{Introduction}
\label{sec:intro}

\IEEEPARstart{T}{he} field of speech separation and speech enhancement has witnessed dramatic improvements in performance with the recent advent of deep learning-based techniques \cite{Weninger2014RNN,Erdogan2015ICASSP04,weninger2015speech, Hershey2016ICASSP03,Isik2016Interspeech09,Yu2017PIT,kolbaek2017uPIT,Wang2018ICASSP04Alternative}.
Most of these algorithms rely on the estimation of some sort of time-frequency (T-F) mask to be applied to the time-frequency representation of an input mixture signal, the estimated signal then being resynthesized using some inverse transform. 
Let us denote by $\vec{X}=(x_{t,f})$, $\vec{S}=(s_{t,f})$, and $\vec{N}=(n_{t,f})$ the complex-valued time-frequency representations of a mixture signal, a target source signal, and an interference signal, respectively, where $t$ denotes the time frame index and $f$ the frequency bin index. We also denote by $\theta_{t,f} = \angle (s_{t,f}/x_{t,f})$ the phase difference between the mixture and the target source. The time-frequency representation is here typically taken to be the short-term Fourier transform (STFT), such that $x_{t,f} = s_{t,f} + n_{t,f}$. The goal of speech enhancement or separation can be formulated as that of recovering an estimate $\hat{\vec{S}}=(\hat{s}_{t,f})$ of $\vec{S}$ from $\vec{X}$, and we're interested in particular in algorithms that do so by estimating a mask $\vec{C}=(c_{t,f})$ such that $\hat{s}_{t,f}=c_{t,f} x_{t,f}$. Note that the interference signal itself could also be a separate target, such as in the case of speaker separation. %

In most cases, these time-frequency masks are real-valued, which means that they only modify the magnitude of the mixture in order to recover the target signal. Their values are also typically constrained to lie between 0 and 1, both for simplicity and because this was found to work well under the assumption that only the magnitude is modified, retaining the mixture phase for resynthesis. 

Several reasons can be cited for focusing on modifying only the magnitude: the noisy phase is actually the minimum mean-squared error (MMSE) estimate \cite{ephraim84} under some simplistic statistical independence assumptions (which typically do not hold in practice); combining the noisy phase with a good estimate of the magnitude is straightforward and gives somewhat satisfactory results; until recently, getting a good estimate of the magnitude was already difficult enough such that optimizing the phase estimate was not a priority, or to put it in other words, phase was not the limiting factor in performance; estimating the phase of the target signal is believed to be a hard problem.

With the advent of recent deep learning algorithms, the quality of the magnitude estimates has improved significantly, to the point that the noisy phase has now become a limiting factor to the overall performance. Because the noisy phase is typically inconsistent with the estimated magnitude \cite{LeRoux2008SAPA09phase,Gerkmann2015SPM03}, the reconstructed time-domain signal has a different magnitude spectrogram from the intended, estimated one.
As an added drawback, further improving the magnitude estimate by making it closer to the true target magnitude may actually lead to worse results when pairing it with the noisy phase, in terms of performance measures such as signal to noise ratio (SNR). Indeed, if the noisy phase is incorrect and for example opposite to the true phase, using $0$ as the estimate for the magnitude is a ``better'' choice than using the correct magnitude value, which may point far away in the wrong direction. Using the noisy phase is thus not only sub-optimal as a phase estimate, it likely also forces the magnitude estimation algorithms to limit their accuracy with respect to the true magnitude.

\begin{figure}[t]
	\centering
	\includegraphics[width=.6\columnwidth]{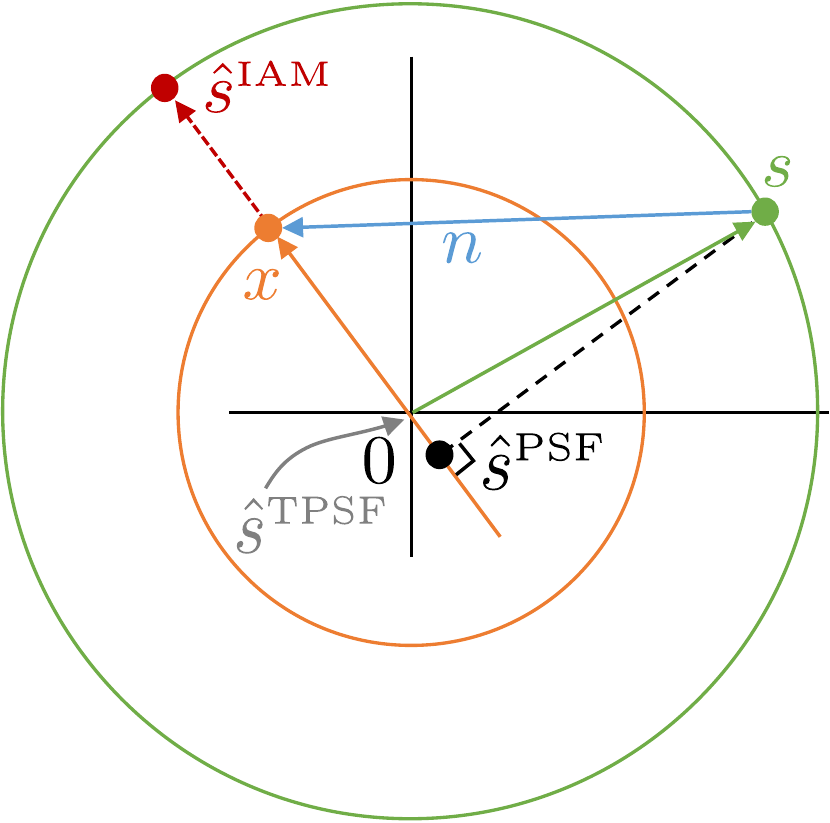}
	\caption{Illustration of the complex mask estimates obtained when using the noisy/mixture phase. The closest point to the clean source $s$ along the line of estimates with phase equal to that of the mixture $x$ is $\hat{s}^{\text{PSF}}$, whose magnitude is very different from the true clean magnitude. The point along that line with true clean magnitude $\hat{s}^{\text{IAM}}$ lies further from the clean source $s$.}
	\label{fig:noisy_phase_issues}\vspace{-.3cm}
\end{figure}

Together with the simplicity of using logistic sigmoid output activations, use of the mixture phase is in particular one of the reasons why mask estimation algorithms typically do not attempt to estimate mask values larger than $1$. %
Indeed, such values are expected to occur in regions where there was a canceling interference between the sources, and it is likely that the noisy phase is a bad estimate there; increasing the magnitude without fixing the phase is thus likely to bring the estimate further away from the target, compared to where the original mixture was in the first place. These issues are illustrated in Fig.~\ref{fig:noisy_phase_issues}, where for simplicity we only consider the case of a single T-F bin in the complex plane, and we omit the time-frequency subscripts $t,f$. The phase-sensitive filter (PSF) estimate $\hat{s}^{\text{PSF}} = \cos(\theta) \frac{|s|}{|x|} x $ corresponds to the orthogonal projection of the clean source $s$ on the line defined by the mixture $x$ \cite{Erdogan2015ICASSP04}; because of the cancelling interference, the PSF estimate here lies in the opposite direction of the mixture. The truncated PSF estimate $\hat{s}^{\text{TPSF}}$, where the mask is constrained to lie in $[0,1]$, is thus equal to $0$ here. The ideal amplitude mask (IAM) estimate $\hat{s}^{\mathrm{IAM}} = \frac{|s|}{|x|}x$, which has the correct clean magnitude, is further from the clean source than either $0$ or the PSF estimate.

\begin{figure}[t]
	\centering
	\includegraphics[width=.99\columnwidth]{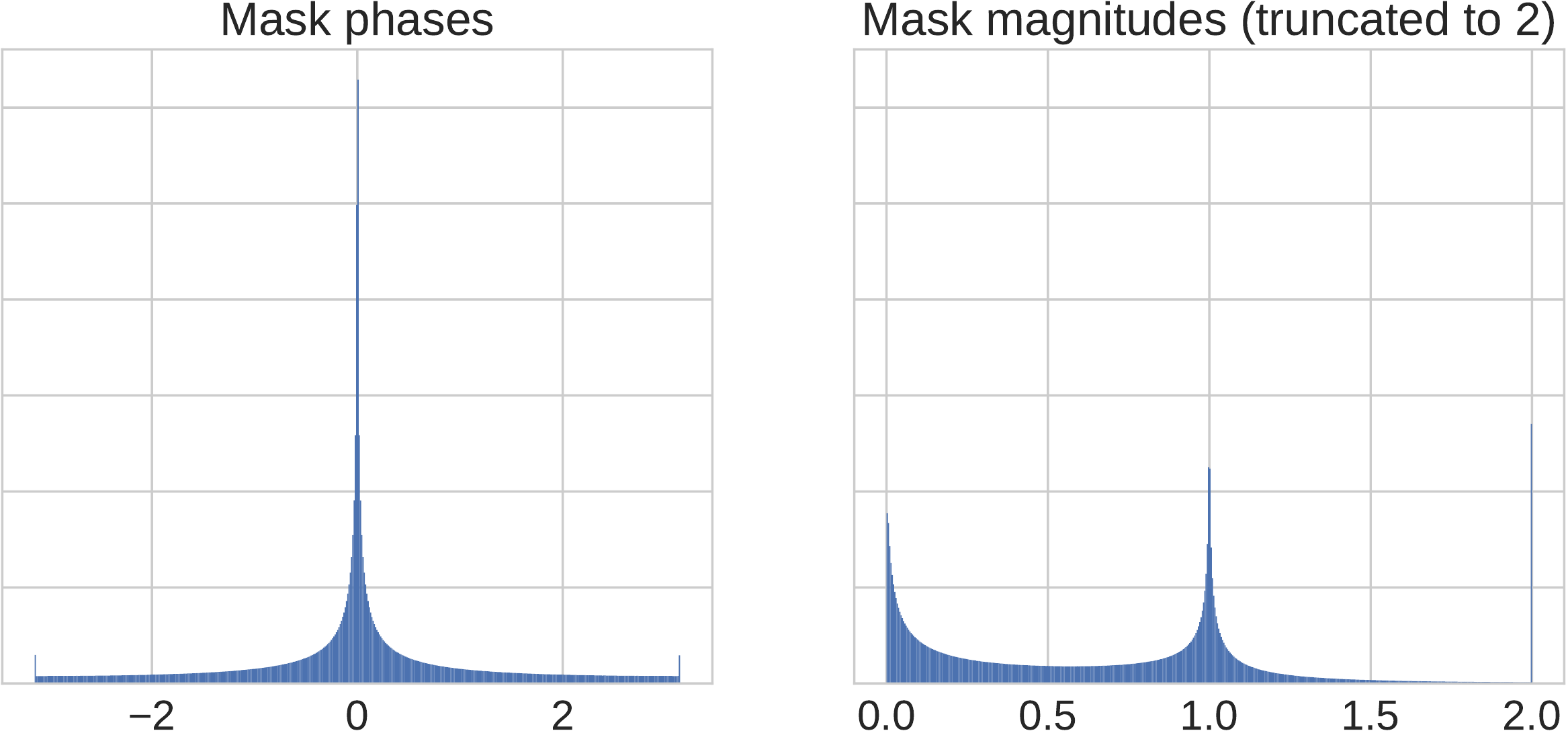}
	\caption{Phase and magnitude distributions of $s_{t,f}\big/x_{t,f}$, truncated to $R_{\mathrm{max}}$}
	\label{fig:mask_statistics}
\end{figure}

By improving upon the noisy phase, we could thus unshackle magnitude estimation algorithms and allow them to attempt bolder estimates that are also more faithful to the true reference magnitude, unlocking new heights in performance. In particular, it would now be worth attempting to involve mask estimates that are allowed to go beyond $1$. For example, one may consider estimating the IAM mentioned above, 
or a version of it truncated to $[0,R_\mathrm{max}]$.
One may also consider estimating a discretized magnitude mask, where the discrete values are not restricted to lie in $[0,1]$. An example of typical distributions for the magnitude and phase components of the ideal complex mask $a^{\mathrm{ICM}}_{t,f}=\frac{s_{t,f}}{x_{t,f}} = a^{\mathrm{IAM}}_{t,f} \mathrm{e}^{\j \theta_{t,f}} $, with the magnitude truncated to $R_\mathrm{max} = 2$, are shown in Fig.~\ref{fig:mask_statistics}. It is clear that a significant proportion of the magnitude mask data lies strictly above $1$.

We have already started exploring this scheme for the magnitude, with the introduction of a convex softmax activation function which interpolates between the values $0,1,2$ to obtain a continuous representation of the interval $[0,2]$ as the target interval for the magnitude mask \cite{Wang2018Interspeech09}. We showed that this activation function led to significantly better performance when optimizing for best reconstruction after a phase reconstruction algorithm. This intuitively makes sense, because the reconstructed phase used to obtain the final time-domain signal is likely to better exploit a magnitude estimate more faithful to the clean magnitude, in particular at time-frequency bins where the clean magnitude is larger than the mixture magnitude due to cancelling interference.

We propose here a generalization of this idea of relying on discrete values to build representations for the masks. We extend the concept of convex softmax activation for the magnitude to the combination of a magnitude codebook, or magbook, with a softmax layer to build various magnitude representations, either discrete or continuous. Similarly, we propose to combine a phase codebook, or phasebook, with a softmax layer to build various phase representations, again either discrete or continuous. Finally, we propose an alternate representation which foregoes the factorization between magnitude and phase and combines a complex codebook, or combook, with a softmax layer to build various complex mask representations.
These representations are flexible and can be incorporated within optimization frameworks that are regression-based, classification-based, or a combination of both.

{\bf Related works:} This paper's contributions are at the intersection of multiple directions of research: classification-based separation, discrete phase representations, complex mask estimation, phase-difference modelling, and phase reconstruction.
The idea of considering separation as a classification problem was explored first using shallow methods, in particular support vector machines \cite{wang2005ideal,Hu2006monaural,kim2009algorithm}, and later deep neural networks \cite{Wang2012cocktail}, and was arguably at the onset of the deep learning revolution in this field.
A few works have proposed to consider discrete representations of the phase for source separation, such as \cite{rennie2005variational} and  \cite{Liutkus2018ICASSP04}, in both cases within a generative model based on mixtures of Gaussians.
Some works have attempted to incorporate phase modeling for deep-learning-based source separation, in particular with the so-called complex ratio mask
\cite{williamson2016complex}, which does consider ranges of values that are not limited to $[0,1]$. While the complex ratio mask used a continuous real-imaginary representation, we here focus mainly on discrete representations involving a magnitude-phase factorization or a direct modelling of the complex value (with the real and imaginary parts considered jointly). We also model not the clean phase but a phase mask, that is, a phase difference between the mixture and the clean source, or in other words a correction to be applied to the mixture phase to get closer to the clean phase. Estimating the phase difference was recently considered within an audio-visual separation framework in \cite{afouras2018conversation}, where it is reconstructed using a convolutional network that takes the estimated magnitude and the noisy phase as input.
Another, potentially complementary, way to improve the phase is to use phase reconstruction. Recent works from our team applied phase reconstruction at the output of a good magnitude estimation network \cite{Wang2018ICASSP04Alternative}, then trained through an unfolded iterative phase reconstruction algorithm \cite{Wang2018Interspeech09}. We finally trained the time-frequency representations used within the phase reconstruction algorithm themselves \cite{Wichern2018IWAENC09}, which is the current state-of-the-art in methods relying on time-frequency representations.

As we were finalizing this article, two related works worth mentioning were published. First, a deep-learning-based source separation algorithm, referred to as PhaseNet \cite{takahashi2018phasenet}, attempts to estimate discretized values of the target source phase; the discretized values are fixed to a uniform quantization along the unit circle, and the network is trained using cross-entropy. As it will become clear in this article, apart from the fact that PhaseNet attempts to estimate the target phase instead of the phase difference, the representation used corresponds to a particular setup of our framework, with a fixed uniform phasebook, cross-entropy training, and argmax based inference. Our framework allows for much more variety in both training and inference schemes, in particular allowing fully end-to-end training which is cumbersome with argmax inference. Second, an updated version of the TasNet algorithm \cite{Luo2018TasNet09arXiv} just established a new state-of-the-art on the wsj0-2mix dataset, surpassing our previous numbers as well as those presented in this article. The TasNet article introduced several interesting techniques that could be adopted in our framework, such as the use of convolution layers instead of recurrent ones, layer normalization schemes, and the use of SI-SDR as the objective instead of the $L^1$ waveform approximation loss that we consider. It is unclear how much these techniques would influence the performance of TasNet's competing methods, and we shall consider incorporating them in our framework as future work.

\section{Designing masks based on discrete representations}

We propose to rely on discrete values to build representations for a complex ratio mask, either via its factorization into magnitude and phase components or directly as a complex value. In particular, we propose to model the magnitude mask using a combination of a magnitude codebook, or magbook, with a softmax layer, and to model the phase mask (i.e., the correction term between mixture phase and clean phase) using a combination of a phase codebook, or phasebook, with a softmax layer. Alternatively, we consider modelling the complex ratio mask directly %
using a combination of a complex codebook, or combook, with a softmax layer; magnitude and phase are then modelled jointly.

\begin{figure}[t]
	\centering
		\includegraphics[width=.6\columnwidth]{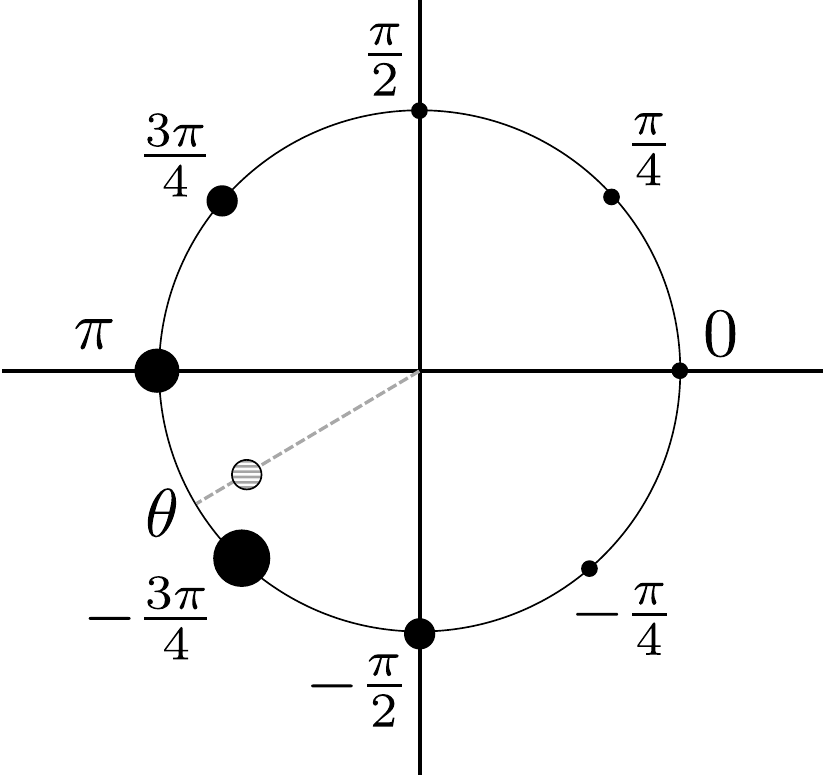}
	\caption{Illustration of the phase interpolation scheme for a uniform phasebook with 8 elements. Softmax probabilities are displayed via the surface of each circle.}
	\label{fig:phase_interpolation}
\end{figure}

We consider scalar codebooks  $\mathcal{M}_M=\{m^{(1)},\dots,m^{(M)}\}$ for the magnitude mask,  $\mathcal{F}_P=\{\theta^{(1)},\dots,\theta^{(P)}\}$ for the phase mask, and $\mathcal{C}_C=\{c^{(1)},\dots,c^{(C)}\}$ for the complex mask. At each time-frequency bin $t,f$, a network can estimate softmax probability vectors for the magnitude mask, the phase mask, or the complex mask, denoted by 
\begin{align}
p_{\bm{\phi}}(m_{t,f} | \vec{O}) &\in \Delta^{M-1},\\ p_{\bm{\phi}}(\theta_{t,f} | \vec{O}) &\in \Delta^{P-1},\\ p_{\bm{\phi}}(c_{t,f} | \vec{O}) &\in \Delta^{C-1},
\end{align}
where $\vec{O}$ denotes the input features, $\bm{\phi}$ the network parameters, and $\Delta^{n}= \left\{(t_0,\dots,t_n)\in\mathbb{R}^{n+1}\mid\sum_{i = 0}^{n}{t_i} = 1 \mbox{ and } t_i \ge 0 \mbox{ for all } i\right\}$ is the unit $n$-simplex. %
We consider several options for using these softmax layer output vectors in order to build a final output, either as probabilities, to sample a value or to select the most likely one, or as weights within some interpolation scheme:
\begin{itemize} %
	\item select the one-best (``argmax''): 
	\begin{align}
		m^{\text{out}}_{t,f} &= \argmax p_{\bm{\phi}}(m_{t,f} | \vec{O})\\
		\theta^{\text{out}}_{t,f} &= \argmax p_{\bm{\phi}}(\theta_{t,f} | \vec{O}) \label{eq:phase_argmax}\\		c^{\text{out}}_{t,f} &= \argmax p_{\bm{\phi}}(c_{t,f} | \vec{O})
	\end{align}
	\item sample from the softmax distribution (``sampling''): 
	\begin{align}
		m^{\text{out}}_{t,f} &\sim  p_{\bm{\phi}}(m_{t,f} | \vec{O})\\
		\theta^{\text{out}}_{t,f} &\sim  p_{\bm{\phi}}(\theta_{t,f} | \vec{O})\\
		c^{\text{out}}_{t,f} &\sim  p_{\bm{\phi}}(c_{t,f} | \vec{O})
	\end{align}
	\item compute the expected value over the distribution (``interpolation''): 
	\begin{align}
		m^{\text{out}}_{t,f} &= \sum_{i} p_{\bm{\phi}}(m_{t,f} = m^{(i)} | \vec{O})\, m^{(i)}\\
		\theta^{\text{out}}_{t,f} &= \angle \sum_{j} p_{\bm{\phi}}(\theta_{t,f} = \theta^{(j)}| \vec{O})\, \mathrm{e}^{\j \theta^{(j)}} \label{eq:phase_interp}\\
		c^{\text{out}}_{t,f} &= \sum_{k} p_{\bm{\phi}}(c_{t,f} = c^{(k)} | \vec{O})\, c^{(k)}.
	\end{align}		
\end{itemize}
Note that the interpolation for the phase in Eq.~\eqref{eq:phase_interp} is performed in the complex domain and that taking the angle implies a renormalization step; this interpolation is illustrated in Fig.~\ref{fig:phase_interpolation}. Further note that the interpolation scheme for the magnitude is an extension of the classical sigmoid activation function for the case of a fixed magbook of size $2$ with elements $\{0,1\}$ (referred to here as uniform magbook 2), and an extension of the convex softmax considered in \cite{Wang2018Interspeech09} for the case of a fixed magbook of size $3$ with elements $\{0,1,2\}$ (referred to here as uniform magbook 3).

In the following, we shall call ``phasebook layer'' a layer computing phase values based on the outputs of a softmax layer and a phasebook via a method such as those above, and similarly for a ``magbook layer'' and a ``combook layer''.

There are multiple motivations for using such representations.
For both magnitude and phase, the combination of a discrete codebook with a softmax layer leads to a very flexible framework, where one can define both discrete and continuous representations which can be involved in both classification-based and regression-based optimization frameworks. The continuous representations may lead to more accurate estimates, or be easier to include within an end-to-end training scheme. On the other hand, the discrete representations open the possibility to consider conditional probability relationships across variables combined with the chain rule, and may also avoid regression issues, for example where the estimated value is an interpolation of two values with high probability but itself has low probability. For the magnitude specifically, as mentioned above, this representation provides a way to generalize classical activations.
For the phase specifically, relying on discrete values makes it possible to design simple representations that take into account phase wrapping, that is, the fact that any measure of difference between phase values should be considered modulo $2\pi$. Indeed, if the phasebook values are used as is, either via sampling or $\mathrm{argmax}$ selection, there is no need to introduce a notion of proximity between various values; if the phasebook values are used within an interpolation scheme in the complex domain such as in Eq.~\eqref{eq:phase_interp},
then the phase is defined by its location around the unit circle,  varies continuously with the softmax probabilities, and values such as $-\pi+\epsilon$ and $\pi-\epsilon$ for small $\epsilon$ can be obtained with softmax probabilities that are close to each other. This would not be the case if one for example modelled phase via a linear transformation of a logistic sigmoid function, such as $\pi+2\pi\sigma(u), u\in \RR$: then, $-\pi+\epsilon$ and $\pi-\epsilon$ would be represented internally by the network via values very far from each other. Regarding phase, note that one could use the same representation to directly model the clean phase instead of a phase difference, or in addition to it and then combine the two estimates. %

\section{Phasebook with argmax}

To get an idea of the potential benefits of a better phase modeling, we first consider the argmax scheme for the phase mask, in which the system attempts to select the best codebook value at each T-F bin.

Given a phasebook $\mathcal{F}_P=\{\theta^{(1)},\dots,\theta^{(P)}\}$, the goal of our system is to estimate at each T-F bin $(t,f)$ the codebook index $j_{t,f}$ such that:
\begin{equation}
j_{t,f} = \argmin_{j} | m_{t,f} e^{\j \theta^{(j)}} x_{t,f} - s_{t,f} |^2,
\end{equation}
where $m_{t,f}$ is some estimate for the magnitude of the mask.
The estimation is in fact independent of the magnitude mask value:
\begin{equation}
j_{t,f} = \argmin_{j} \cos (\theta^{(j)} - \angle (s_{t,f} / x_{t,f})).\label{eq:estep-f}
\end{equation}

\subsection{Codebook optimization}

An important question is how to best design the phasebook. An obvious and easy choice is to use regularly spaced values. But ideally, one would like to optimize them for best performance on some training data. This can be done independently of the classification system, or together with it, optimizing both the phasebook and the classification system jointly in an end-to-end fashion. We first consider how to optimize the codebook offline in a pre-training step, for optimal performance given a magnitude estimate. That magnitude estimate may be obtained either with a pre-trained magnitude estimation network, or with an oracle mask.

The objective function for the phasebook training is:
\begin{equation}
\mathcal{S}(\mathcal{F}_P) = \sum_{f,t} \min_{j} \big| m_{t,f} e^{\j f^{(j)}} x_{t,f} - s_{t,f} \big|^2.
\end{equation}
It can be optimized using an EM-like algorithm. %
In the E-step, the optimal codebook assignments are computed for each T-F bin according to Eq.~\ref{eq:estep-f}. %
In the M-step, we update the phasebook to further decrease the objective function
by solving 
\begin{equation}
\argmin_{\theta^{(j_0)}} \sum_{(t,f)| \theta_{t,f}=\theta^{(j_0)}} |x_{t,f}|^2 \Big| m_{t,f} e^{-\j \theta^{(j_0)}} - \frac{s_{t,f}}{x_{t,f}} \Big|^2, \label{eq:mstep-f_obj}
\end{equation}
which can easily be shown to be equivalent to
\begin{equation}
\argmax_{\theta^{(j_0)}} \real \Big( \Big( \sum_{(t,f)| \theta_{t,f}=\theta^{(j_0)}} |x_{t,f}|^2 \frac{s_{t,f}}{x_{t,f}}  m_{t,f} \Big) e^{-\j \theta^{(j_0)}} \Big), \label{eq:mstep-f_obj2}
\end{equation}
leading to the following update equation:
\begin{equation}
\theta^{(j_0)} \leftarrow \angle \Big( \sum_{(t,f)| \theta_{t,f}=\theta^{(j_0)}} m_{t,f} |x_{t,f}|^2 \frac{s_{t,f}}{x_{t,f}}  \Big). \label{eq:mstep-f}
\end{equation}

Note that a magbook could be similarly (and jointly) optimized under an argmax scheme, at each step looping in order over the updates of the magbook values, the magbook assignments, the phasebook assignments, and the phasebook values, the latter two as described above. %
Finally, optimization of a combook under an argmax scheme can be simply obtained via the k-means algorithm.

In our experiments, we optimize the codebooks on a speech separation task using 50 randomly selected utterances from the wsj0-2mix training dataset \cite{Hershey2016ICASSP03}. Note that we noticed similar behaviors in terms of optimized codebook configurations and separation performance on a speech enhancement task with data from the CHiME2 training set \cite{Vincent13-TSC}. The initial codebooks can be randomly sampled from the data, or set manually. In the latter case, the phasebooks are initialized using uniform codebooks with values that partition the unit circle into equal angular intervals, making sure that $0$ is one of the elements of the codebook: 
$\mathcal{F}_P^{\text{uniform}}=\{0,\dots,\frac{2p\pi}{P},\dots, \frac{2(P-1)\pi}{P}\}$.
We run the optimization algorithm for $40$ epochs, which was enough to ensure convergence. It is likely that the output of the optimization is only a local optimum, and even better codebooks could potentially be obtained by running multiple optimizations with different initializations, but we did not consider this here.

\begin{figure}[t]
	\centering
		\includegraphics[width=.99\columnwidth]{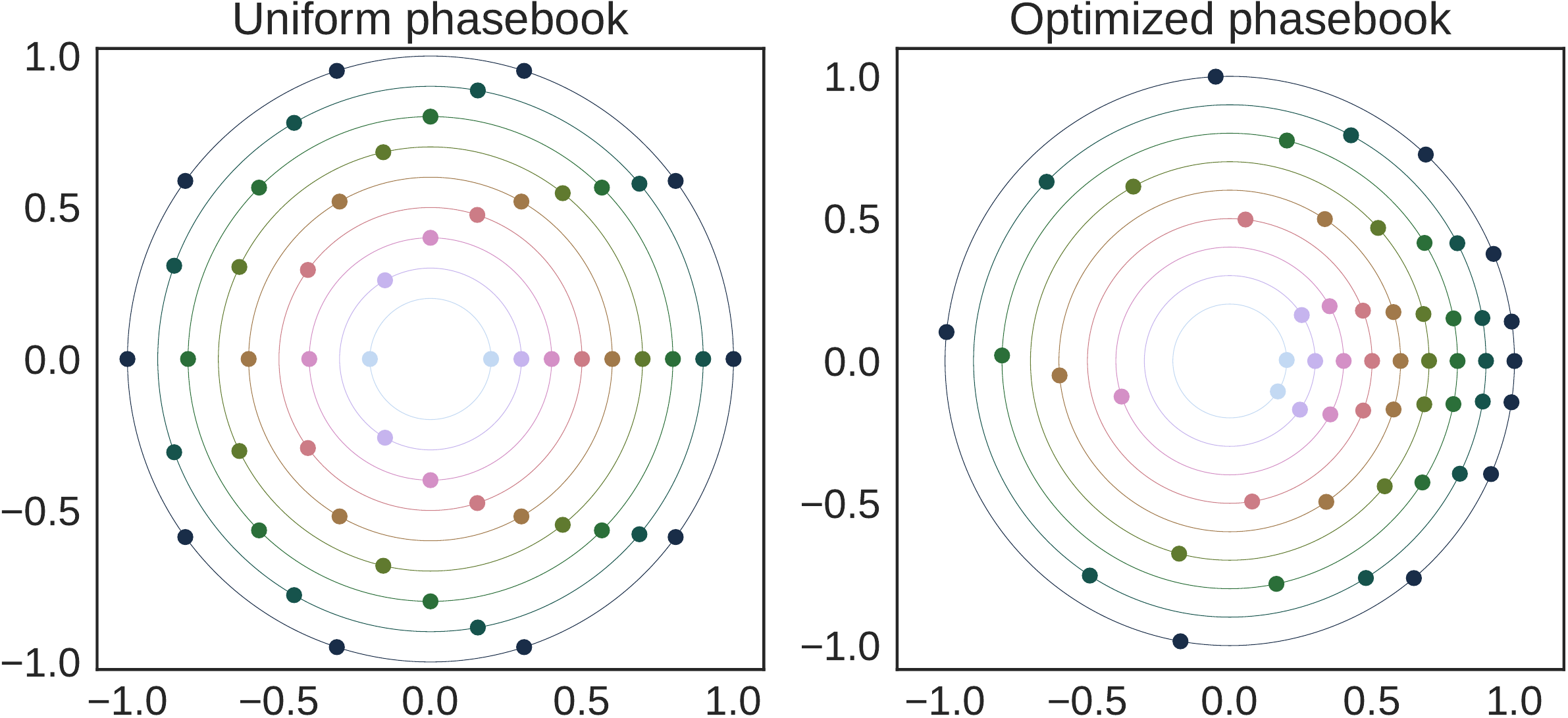}
	\caption{Uniform and optimized phasebooks for $P=2,\dots,10$ and an oracle IAM estimate for magnitude, where the radius of each circle is equal to $P/10$.}
	\label{fig:optimized_phase_codebooks}
\end{figure}

Figure~\ref{fig:optimized_phase_codebooks} shows the optimized phasebooks for $P=2,\dots,10$ and a magnitude obtained using an oracle IAM magnitude mask, together with the uniform phasebooks they were initialized from. %

\subsection{Oracle performance}

\begin{figure}[t]
	\centering
	\includegraphics[width=.99\columnwidth]{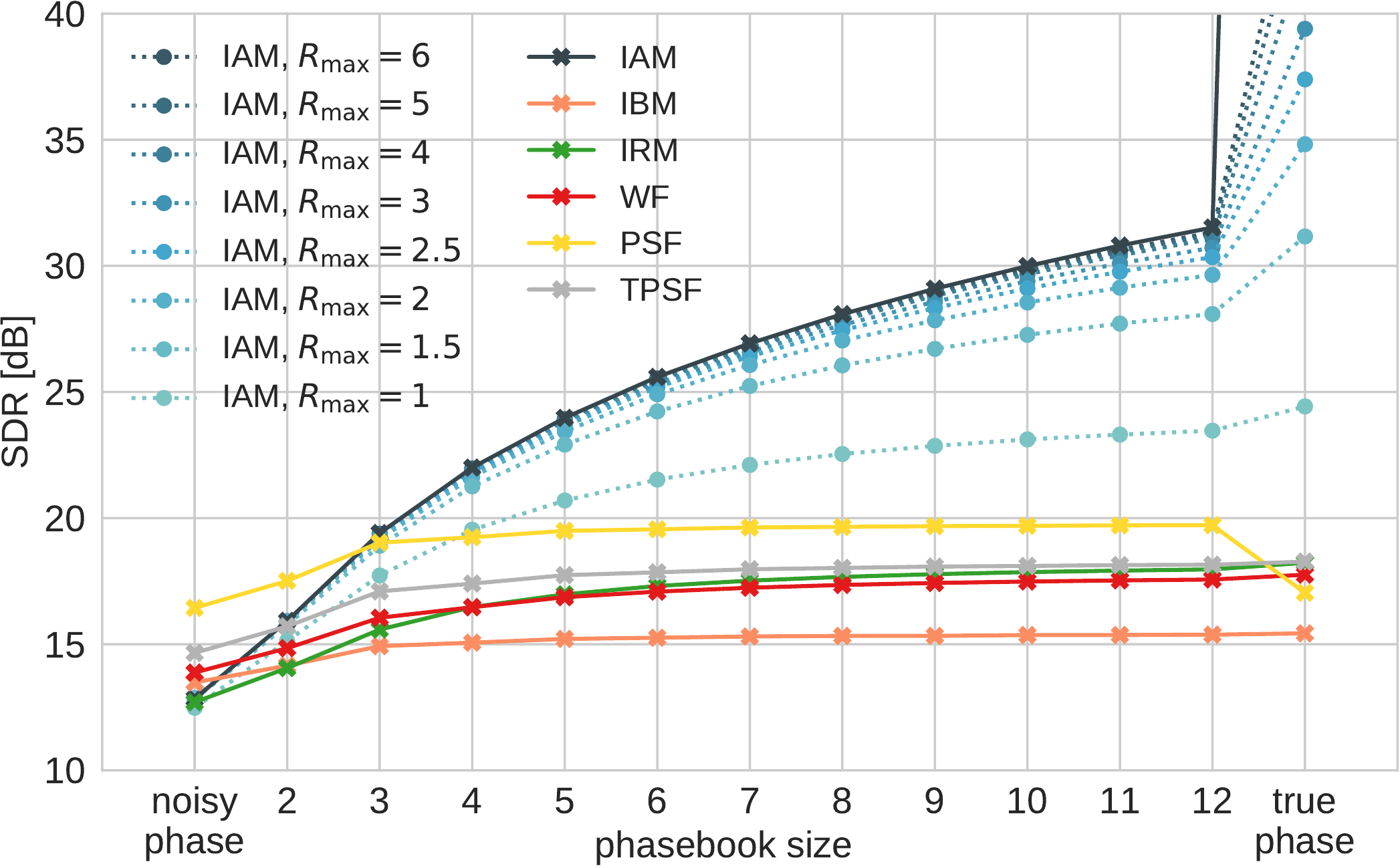}
	\caption{Speech SI-SDR improvement (dB) for truncated ideal amplitude mask and various classical masks with quantized phase difference, using optimal phasebooks of various sizes.}
	\label{fig:mask_comparison}
\end{figure}

We compare here the performance of various classical masks as well as truncated ratio masks with various truncation thresholds $R_{\mathrm{max}}$ in terms of scale-invariant signal-to-distortion ratio (SI-SDR), which we define here as the scale-invariant signal-to-noise ratio between the target speech and the estimate \cite{LeRoux2018SISDR}. The evaluation is performed under oracle conditions (i.e., the mask values are obtained using both the mixture and the true reference signals) on the full wsj0-2mix evaluation set \cite{Hershey2016ICASSP03}. For each mask, we report results where we combined the magnitude part of the mask with the noisy phase, the true phase (i.e., that of the reference), and quantized phases using phasebooks with $P=2,\dots,10$ elements, each phasebook being optimized for the particular magnitude mask it is used with similarly to the algorithm described above. The results are shown in Fig.~\ref{fig:mask_comparison}.

The classical masks we investigate are the most popular types of masks that were reviewed and whose oracle performance when paired with the noisy phase was compared in \cite{Erdogan2015ICASSP04}. They include the ideal amplitude mask (IAM), phase sensitive filter (PSF), and its truncated version to $[0,1]$ (TPSF), all defined in Section~\ref{sec:intro}, as well as the ideal binary mask (IBM: $a^{\mathrm{IBM}} = \delta(|s| > |n|)$), ideal ratio mask (IRM: $a^{\mathrm{IRM}} = |s|\big/ (|s| + |n|) $), and Wiener-filter-like mask (WF: $a^{\mathrm{WF}} = |s|^2 \big/ (|s|^2 + |n|^2)$). All these masks are real-valued, and only modify the magnitude of the mixture signal (except for PSF, which allows a reversal of the phase). 

We first notice that, apart from the phase-sensitive masks PSF and TPSF, all masks lead to similar results when paired with the noisy phase. This confirms that the noisy phase drastically limits the performance.
As soon as a slightly better estimate of the phase is considered, performance significantly increases, especially for those masks that consider magnitude ratio values above $1$. For phases other than the noisy phase, we notice a very big jump in performance when allowing the truncation ratio to go from a classical value $R_\mathrm{max}=1$ to an only slightly larger value $R_\mathrm{max}=1.5$. Interestingly, very small codebook sizes already lead to very high oracle performance, e.g., $P=4$. In non-oracle conditions, of course, we need to find the right balance between upper-bound performance and classification accuracy.

\begin{figure}[t]
	\centering
	\includegraphics[width=.99\columnwidth]{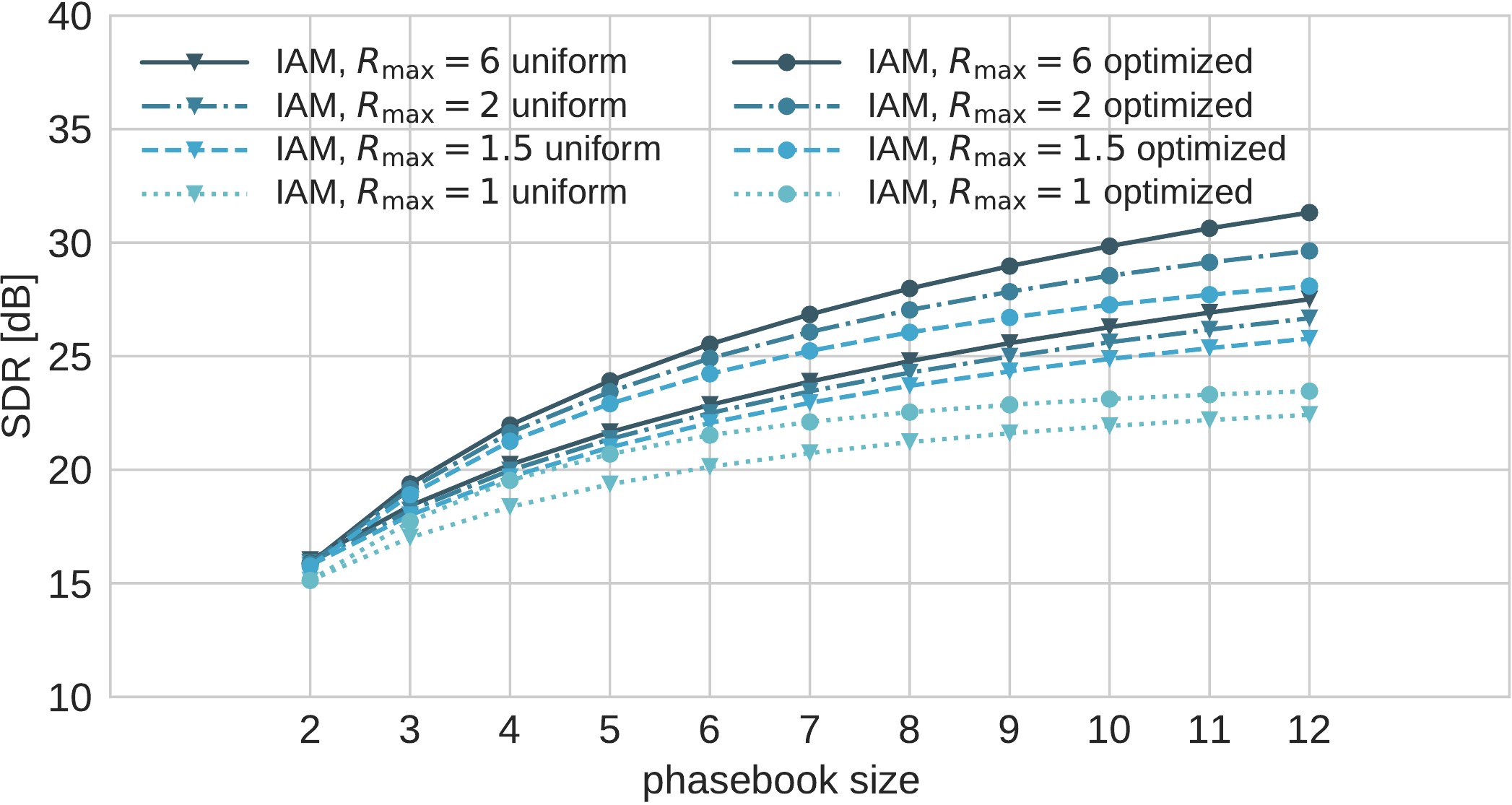}
	\caption{Influence of codebook optimization for truncated ideal amplitude masks with quantized phase.}
	\label{fig:opt_vs_reg}
\end{figure}

Fig.~\ref{fig:opt_vs_reg} shows results with uniform and optimized phasebooks for truncated ideal amplitude masks. Optimizing the codebooks leads in all cases to significant improvements, with typical gains around 2 to 3~dB. %

\section{Objective functions}
\label{sec:objectives}

We consider the above representations as layers within a deep learning model for source separation, and we need to optimize the parameters $\bm{\phi}$ of the model under some objective function. We note that the magbook $\mathcal{M}_M$, phasebook $\mathcal{F}_P$, and combook $\mathcal{C}_C$ themselves can be considered fixed (to uniform or pre-trained values as described in the previous section), or optimized jointly with the rest of the network, with the codebook values considered as part of the network parameters.

We present multiple objective functions for the magnitude and phase components as well as for the complex mask; in practice, these objective functions can be combined with each other within a multi-task learning framework. Note also that, for simplicity, we define here the objective functions on a single source-estimate pair, but the definitions can be straightforwardly extended to the permutation-free training scheme commonly used in speech separation \cite{Hershey2016ICASSP03,Isik2016Interspeech09,Yu2017PIT}.

\subsection{Cross-entropy objectives}

Let $\vec{i}^{\mathrm{ref}}$ denote the reference values for the magnitude mask, and $\bm{j}^{\mathrm{ref}}$ the reference values for the phase mask, which are here the corresponding reference codebook indices. The reference indices for the phase can be obtained using Eq.~\eqref{eq:estep-f}. The reference indices for the magnitude depend on the phase mask that is expected to be used, for example a true reference phase mask as defined above or a current estimate obtained by a network. For a phase mask value $\theta_{t,f}$ at bin $t,f$, the corresponding optimal mask magnitude index is obtained as
\begin{equation}
i_{t,f}^{\mathrm{ref}} = \argmin_{i} | m^{(i)} e^{\j \theta_{t,f}}x_{t,f} - s_{t,f} |,
\end{equation}
or equivalently
\begin{equation}
i_{t,f}^{\mathrm{ref}} = \argmin_{i} \Big| m^{(i)} - \real\big( \frac{s_{t,f}}{x_{t,f}} e^{-\j \theta_{t,f}}\big) \Big|. \label{eq:estep-m}
\end{equation}
The reference indices for the complex mask are denoted as $\vec{k}^{\mathrm{ref}}$ and simply obtained for each T-F bin as the index of the complex number in the codebook that is closest to the ratio mask $\frac{s_{t,f}}{x_{t,f}}$ for some distance, for example $L^2$.

We can now define an objective function based on the cross-entropy against the oracle codebook assignments for the softmax layer outputs of the magbook, phasebook, and combook layers respectively as:
\begin{align}
\hspace{-.25cm}\mathcal{L}_\text{CE-mag}(\bm{\phi}) %
&= - \sum_{t,f}\sum_i \delta(i,i_{t,f}^\mathrm{ref}) \log p_{\bm{\phi}}(m_{t,f} = m^{(i)} | \vec{O}), \\
\hspace{-.25cm}\mathcal{L}_\text{CE-phase} (\bm{\phi}) %
&= - \sum_{t,f} \sum_j \delta(j,j_{t,f}^\mathrm{ref}) \log p_{\bm{\phi}}(\theta_{t,f}=\theta^{(j)} | \vec{O}),\\
\hspace{-.25cm}\mathcal{L}_\text{CE-com} (\bm{\phi}) %
&= - \sum_{t,f} \sum_k \delta(k,k_{t,f}^\mathrm{ref}) \log p_{\bm{\phi}}(c_{t,f}=c^{(k)} | \vec{O}).
\end{align}

If cross-entropy is used for the magnitude, the phase mask used to compute the reference magnitude can either be fixed (to $0$, to a reference computed offline given some phasebook values, or to an initial estimate obtained by an initial phasebook network), or updated throughout training (using the reference phase mask obtained with the current phasebook if it is being optimized as well, or with the current estimate of the phase mask obtained by the network).

\subsection{Magnitude objectives in the T-F domain: MA, MSA, PSA}
All the classical objectives used to train mask inference networks that modify the magnitude can be used here, such as mask approximation (MA), magnitude spectrum approximation (MSA), and phase-sensitive spectrum approximation (PSA). Any norm can be considered to define these objective functions, with $L^1$ and (squared) $L^2$ being most commonly used.
Using $L^1$ as an example, we can define:
\begin{align}
\hspace{-.45cm}\mathcal{L}_{\text{MA},L^1}(\bm{\phi})%
&= \sum_{t,f} | m^{\text{out}}_{t,f} -  m_{t,f}^\mathrm{ref} | ,\\
\hspace{-.45cm}\mathcal{L}_{\text{MSA},L^1}(\bm{\phi})
&= \sum_{t,f} \big|  m^{\text{out}}_{t,f} |x_{t,f}| - |s_{t,f}| \big| ,\\ %
\hspace{-.45cm}\mathcal{L}_{\text{PSA},L^1}(\bm{\phi})
&= \sum_{t,f} \big|  m^{\text{out}}_{t,f} |x_{t,f}| - |s_{t,f}|\cos(\theta_{t,f}^\mathrm{ref}) \big|,
\end{align}
where $\theta_{t,f}^\mathrm{ref}$ is the oracle phase difference between mixture and target, and $m_{t,f}^\mathrm{ref}$ is an oracle magnitude mask such as the IAM.

\iftrue
\subsection{Complex objectives in the T-F domain: CMA, CSA}
\label{sec:CSA}

We can extend the classical MA and MSA objective functions to complex versions involving the estimated complex mask $\vec{c}^{\text{out}}$, either obtained directly using a combook representation, or obtained by combining magnitude and phase estimates at each T-F bin as 
\begin{equation}
c_{t,f}^{\text{out}}=m_{t,f}^{\text{out}} \mathrm{e}^{\j \theta_{t,f}^{\text{out}}}.
\end{equation}
Again using $L^1$ as an example, we can define a complex mask approximation (CMA) objective using the distance between the reconstructed complex ratio mask $c_{t,f}^{\text{out}}$ and a reference complex ratio mask $c_{t,f}^{\text{ref}}$ (e.g., $c_{t,f}^{\text{ref}}=s_{t,f}/x_{t,f}$):
\begin{equation}
\mathcal{L}_{\text{CMA},L^1}(\bm{\phi})%
= \sum_{t,f} | c_{t,f}^{\text{out}} -  c_{t,f}^{\text{ref}} |.
\end{equation}
We can also define a complex spectrum approximation (CSA) objective using the distance between the reconstructed (i.e., masked) T-F representation and the target T-F representation:
\begin{equation}
\mathcal{L}_{\text{CSA},L^1}(\bm{\phi})%
= \sum_{t,f} | c_{t,f}^{\text{out}} x_{t,f} - s_{t,f}  |.
\end{equation}
	
\fi

\subsection{Time-domain objectives: WA, WA-MISI}

Recently, we introduced a waveform approximation (WA) objective defined on the time-domain signal $\hat{s}[l]$ reconstructed by inverse STFT from the masked mixture  \cite{Wang2018Interspeech09}.
We also proposed training through an unfolded phase reconstruction algorithm such as multiple input spectrogram inversion (MISI) \cite{Gunawan2010}, using the WA objective on the reconstructed time-domain signal $\hat{s}^{(K)}[l]$ after $K$ iterations.

Denoting by $s[l]$ the reference time-domain signal, %
and again using $L^1$ as an example, we define:
\begin{align}
\hspace{-.45cm}\mathcal{L}_{\text{WA},L^1}(\bm{\phi})%
&= \sum_{l} | \hat{s}[l]-s[l]| ,\\
\hspace{-.45cm}\mathcal{L}_{\text{WA-MISI-K},L^1}(\bm{\phi})%
&= \sum_{l} | \hat{s}^{(K)}[l]-s[l]| .
\end{align}
In the same way as we did for magnitude-only mask inference networks \cite{Wang2018Interspeech09}, we can train a network that estimates both a magnitude mask and a phase mask, or alternatively a complex mask, end-to-end using the above time-domain objective functions.

\subsection{Inference considerations and expected loss}
\label{sec:expected_loss}

When using the cross-entropy training objectives, there is no inference scheme to be explicitly selected at training time, as the optimization is performed solely on the softmax outputs. While any of the inference schemes could be used at test time, either argmax or sampling inference seem most appropriate given the discrete characteristics of the cross-entropy objective.

For the objectives defined on the value of the estimated mask, the reconstructed time-frequency representation, or the reconstructed time-domain signal, one does need to select at training time an inference scheme used to obtain the masks, and a natural choice at test time is to use the same inference scheme as the one used during training. The interpolation scheme is by far the most convenient, because it ensures that the objective function is differentiable with respect to all parameters, and the gradients can be easily computed using straightforward back-propagation. The sampling and argmax schemes may also be considered, and would be particularly relevant if we were to introduce conditional-probability  relationships between T-F bins. However, these schemes raise significant difficulties for the optimization, as both sampling-based and argmax-based selection operations break the differentiation chain.

In order to keep a discrete selection step in the training pipeline for a given loss function, one possibility is 
to define a corresponding {\it expected loss} function which considers all possible choices of values in the codebooks in turn to compute a loss term, weighted by their softmax probability. This corresponds to what one would obtain by sampling many times from the softmax outputs and averaging the loss obtained with the corresponding output. For example, the expected loss version of the CSA loss for the magbook-phasebook case can be defined as
\begin{multline}
	\hspace{-.2cm}\text{LS}_{\text{eCSA},L^1}(\bm{\phi}) %
	\!=\! \sum_{t,f} \sum_{i}\sum_{j}
	p_{\bm{\phi}}(m_{t,f}=m^{(i)})
	p_{\bm{\phi}}(\theta_{t,f}=\theta^{(j)})\\
	| m^{(i)} \mathrm{e}^{\j \theta^{(j)}} x_{t,f} - s_{t,f}|. 
\end{multline}
While the sum can in this case be computed exactly by marginalizing over all T-F bins independently, computing such an expectation becomes much trickier for objective functions that include a coupling between the T-F bins. Such is the case for the WA objective, which is defined in the time domain on the inverted T-F representation: the final output depends on all T-F bins, which thus cannot be marginalized over independently, leading to a combinatorial explosion. We could consider approximating the expected loss 
as the sum of the WA losses for a given number of T-F representations obtained by sampling all T-F bins.  
Back-propagation could then be performed using the policy gradient technique in the REINFORCE algorithm \cite{williams1992simple}, similarly to what was done for automatic speech recognition in \cite{hori2018cycle}. Another option would be to rely on the Gumbel-Softmax trick \cite{jang2016categorical,maddison2016concrete}.

Given preliminary results described below on the CSA objective under-performing the WA objective, the significant complexity involved in implementing an expected loss for the WA objective, and the fact that relying on discrete selection instead of interpolation is expected to mostly become relevant when conditional-probability relationships between T-F bins are considered, we leave this line of research for future works.

\section{Experimental validation}

\subsection{Experimental setup}
We validate the proposed algorithms on the publicly available wsj0-2mix corpus \cite{Hershey2016ICASSP03}, which is widely used in speaker-independent speech separation works. It contains 20,000, 5,000 and 3,000 instantaneous two-speaker mixtures in its 30~h training, 10~h validation, and 5~h test sets, respectively. The speakers in the validation set are seen during training, while the speakers in the test set are completely unseen. The sampling rate is 8~kHz.

For our neural networks, we follow the same basic architecture as in \cite{Wang2018Interspeech09}, containing four BLSTM layers, each with 600 units in each direction, followed by output layers. A dropout of $0.3$ is applied on the output of each BLSTM layer except the last one. The networks are trained on 400-frame segments using the Adam algorithm. The window length is 32~ms and the hop size is 8~ms. The square root Hann window is employed as the analysis window and the synthesis window is designed accordingly to achieve perfect reconstruction after overlap-add. A 256-point DFT is performed to extract 129-dimensional log magnitude input features. All systems are implemented using the Chainer deep learning toolkit \cite{tokui2015chainer}.

\subsection{Chimera++ network with phasebook-magbook mask inference head}
\label{sec:phasebook_network}

We build our system based on the state-of-the-art chimera++ network \cite{Wang2018Interspeech09}, which combines within a multi-task learning framework a deep clustering head outputting a $D$-dimensional embedding for each T-F bin ($D=20$ here), and a mask-inference head with convex softmax output which predicts a magnitude mask with values in $[0,2]$. The chimera++ objective function is
\begin{align} \label{eq:chimera}
	\mathcal{L}_{\text{chi}^{++}_\alpha}=\alpha \mathcal{L}_{\text{DC},\text{W}}+(1-\alpha)\mathcal{L}_{\text{MI}}
\end{align} where $\mathcal{L}_{\text{MI}}$ can be any of the objective functions described in Section~\ref{sec:objectives}, and the weight $\alpha$ is typically set to a high value, e.g., 0.975. The loss used on the deep clustering head is the whitened k-means loss 
\begin{align}
    \mathcal{L}_{\text{DC},\text{W}} &=\|V(V^{\T}V)^{-\frac{1}{2}} -  Y(Y^{\T}Y)^{-1}Y^{\T} V (V^{\T}V)^{-\frac{1}{2}} \|_{\F}^2  \nonumber\\
    &=D - \,\text{tr}\big((V^{\T}V)^{-1}V^{\T}Y(Y^{\T}Y)^{-1}Y^{\T}V\big),
\end{align}
where $V\in \RR^{TF\times D}$ is the embedding matrix consisting of vertically stacked embedding vectors, and $Y\in \RR^{TF\times S}$ is the label matrix consisting of vertically stacked one-hot label vector representing which of the $S$ sources in a mixture dominates at each T-F bin.

\begin{figure}[t]
	\centering
		\includegraphics[width=.99\columnwidth]{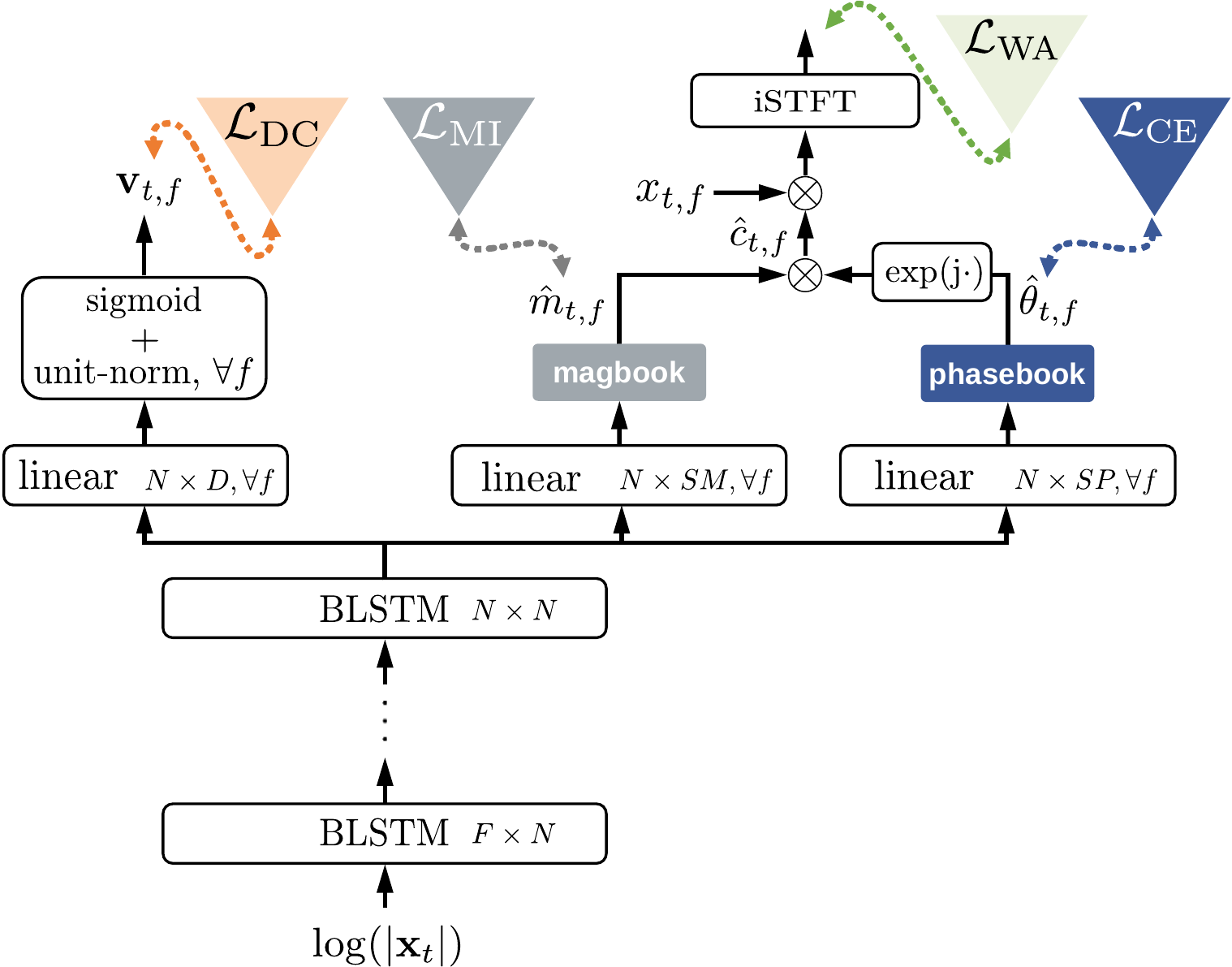}
	\caption{Chimera++ network with phasebook-magbook mask inference head.}
	\label{fig:phasebook-network}
\end{figure}

As we explained above, the mask-inference head with convex softmax output predicting a magnitude mask can be generalized to a magbook layer. We now add a phasebook layer, similar to the magbook layer, as a new head at the output of the final BLSTM layer, as illustrated in Fig.~\ref{fig:phasebook-network}. The final complex mask is obtained by combining the outputs of the magbook and phasebook layers as
\begin{equation}
\hat{c}_{t,f} = \hat{m}_{t,f} \mathrm{e}^{\j \hat{\theta}_{t,f}},
\end{equation}
and then multiplied with the complex mixture to obtain a complex T-F representation $\hat{s}_{t,f}$ of the target estimate:
\begin{equation}
\hat{s}_{t,f} = \hat{c}_{t,f} x_{t,f} = \hat{m}_{t,f} \mathrm{e}^{\j \hat{\theta}_{t,f}} x_{t,f}.
\end{equation}
We still refer to the branch of the network used in computing the final output as the mask-inference (MI) head, which now predicts a complex mask.

\subsection{Training and inference schemes for phasebook}

In this experiment, we start by pre-training chimera++ networks with magbook mask-inference head, where for now we use the fixed convex softmax of \cite{Wang2018Interspeech09} for the magbook layer, referred to here as uniform magbook 3. For each of the MSA, PSA, and WA losses as MI objective function, we train such a network from scratch within the multi-task learning setting involving the deep clustering and MI objectives, then discard the deep clustering head and fine-tune the MI head only. 
\if 0
In preliminary experiments, the CSA loss described in Section~\ref{sec:CSA} was outperformed by the WA loss, so we did not explore it further.
\else 
\fi

We now add a phasebook mask-inference head to these networks as described in Section~\ref{sec:phasebook_network}, where we assume a fixed uniform codebook with values $\frac{2p\pi}{P}, p=0,\dots,P-1$, referred to as uniform phasebook $P$, and
we consider: (1)  training the phasebook layer by itself while keeping the rest of the network fixed, with the cross-entropy loss $\mathcal{L}_\text{CE-phase}$, and using the argmax scheme in Eq.~\ref{eq:phase_argmax} at inference time; (2) training the phasebook layer by itself while keeping the rest of the network fixed, assuming the interpolation scheme in Eq.~\eqref{eq:phase_interp} is used to obtain the final phase mask value, and either the CSA loss $\mathcal{L}_{\text{CSA},L^1}$ or the WA loss $\mathcal{L}_{\text{WA},L^1}$ is used as the training objective; and (3) training the whole network with either the CSA loss $\mathcal{L}_{\text{CSA},L^1}$ or the WA loss $\mathcal{L}_{\text{WA},L^1}$, again assuming the interpolation scheme for the phase.

{\setlength{\tabcolsep}{4pt}
\begin{table}[t]

    \footnotesize
	\centering
	\caption{SI-SDR improvement (dB) on the wsj0-2mix test set for various training paradigms from various pre-trained magnitude estimation networks.} %
	\label{table:ce_e2e}
	\begin{tabular}{lcc|c|c|c}
		\hline\hline
		&\!\!Network\!\!&\!\! Joint mag.\!\!&\multicolumn{3}{c}{Mag.\ pretraining}\\
        Phase estimate &\!\!\!Objective\!\!\!& \!\!training\!\! & MSA & PSA & WA  \\ 
		\hline\hline
		Noisy & - & \batsu & 10.5 & 11.1 & 11.8 \\
		\hline
		Uniform phasebook 8 argmax & CE & \batsu & 10.7 & 11.1 & 11.8\\
		\hline
		Uniform phasebook 8 interp. & CSA & \batsu & 10.5 & 11.1 & 11.8\\
		\hline
		Uniform phasebook 8 interp. & WA & \batsu & 11.2 & 11.1 & 12.0\\
		\hline
        Uniform phasebook 8 interp. & CSA & $\checkmark$ & 11.5 & 11.5 & 11.6\\
		\hline
        Uniform phasebook 8 interp. & WA & $\checkmark$ & 12.2 & 12.4 & 12.4\\
		\hline
		\hline
	\end{tabular}%
\end{table}
}

For this experiment, we consider a uniform phasebook with $P=8$ elements. Results are shown in Table~\ref{table:ce_e2e} in terms of scale-invariant SDR (dB) \cite{LeRoux2018SISDR} on the wsj0-2mix test set.  From Table~\ref{table:ce_e2e}, we see that the CE objective only provides SI-SDR improvements for networks pre-trained with the phase-unaware MSA objective. This intuitively makes sense, as the MSA-based magnitude estimates are likely to be closer to the true magnitude than those obtained with PSA and WA, which try to compensate for the errors in the noisy phase; once the phasebook layer fixes these errors, which it learns to do without considering the interaction with the magnitude in the CE case, the compensation performed by the magnitude estimate may become extraneous or even detrimental. The CSA objective is consistently outperformed by the WA objective both with and without joint training of the magnitude, demonstrating the importance of training through the overlap-add process.  Without joint magnitude training when learning the phasebook layer, the CSA training leads to no difference in SI-SDR, while with the WA objective the largest improvement is again observed for MSA. Finally, when allowing joint training of the magbook layer, all pre-training objectives obtain their best performance, with the exception of the CSA objective with WA pre-training, where  removing the overlap-add process during fine-tuning leads to a performance degradation. Pre-training with PSA and WA obtains slightly larger values than MSA, and overall, the WA objective with the interpolation scheme appears the most robust, both for pretraining and for training networks involving magbook and phasebook layers. We thus focus on this configuration going forward. 

\begin{figure}[t]
	\centering
		\includegraphics[width=.79\columnwidth]{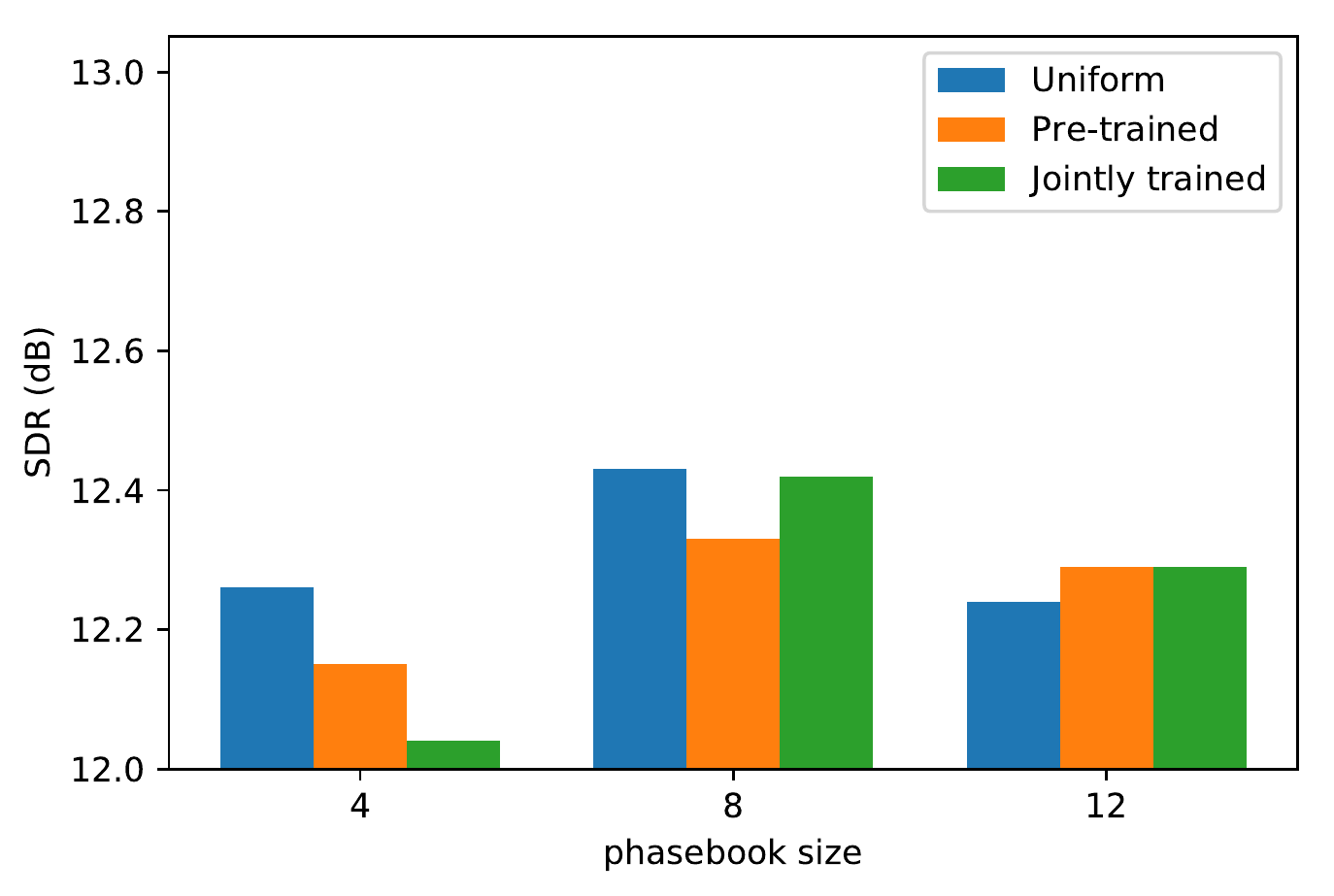}
	\caption{SI-SDR improvement (dB) for various phasebook configurations. }
	\label{fig:bar_sdr_phasebook_type}
\end{figure}

\begin{figure*}[t]
	\centering
		\includegraphics[width=.6\columnwidth]{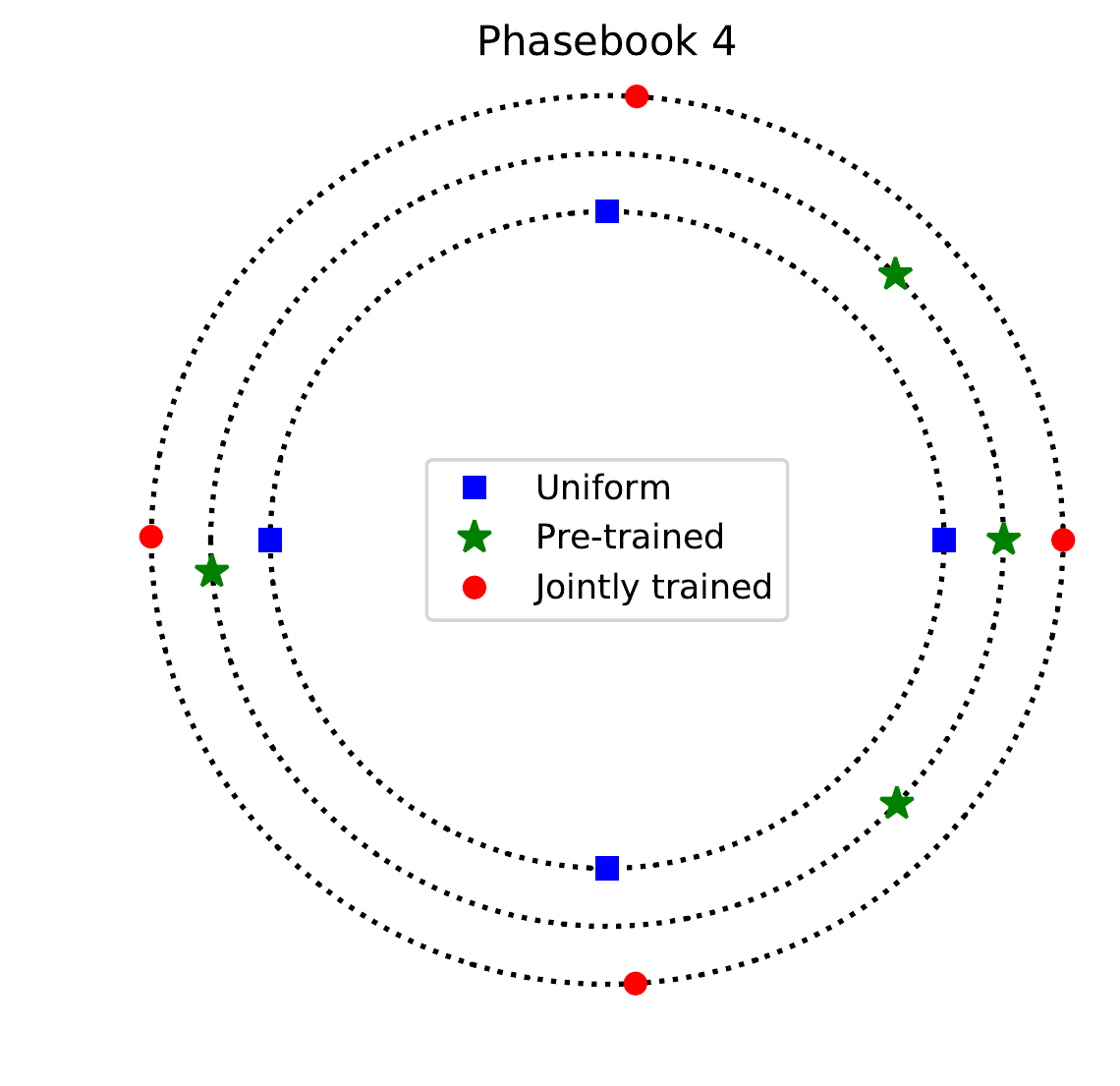}
		\includegraphics[width=.6\columnwidth]{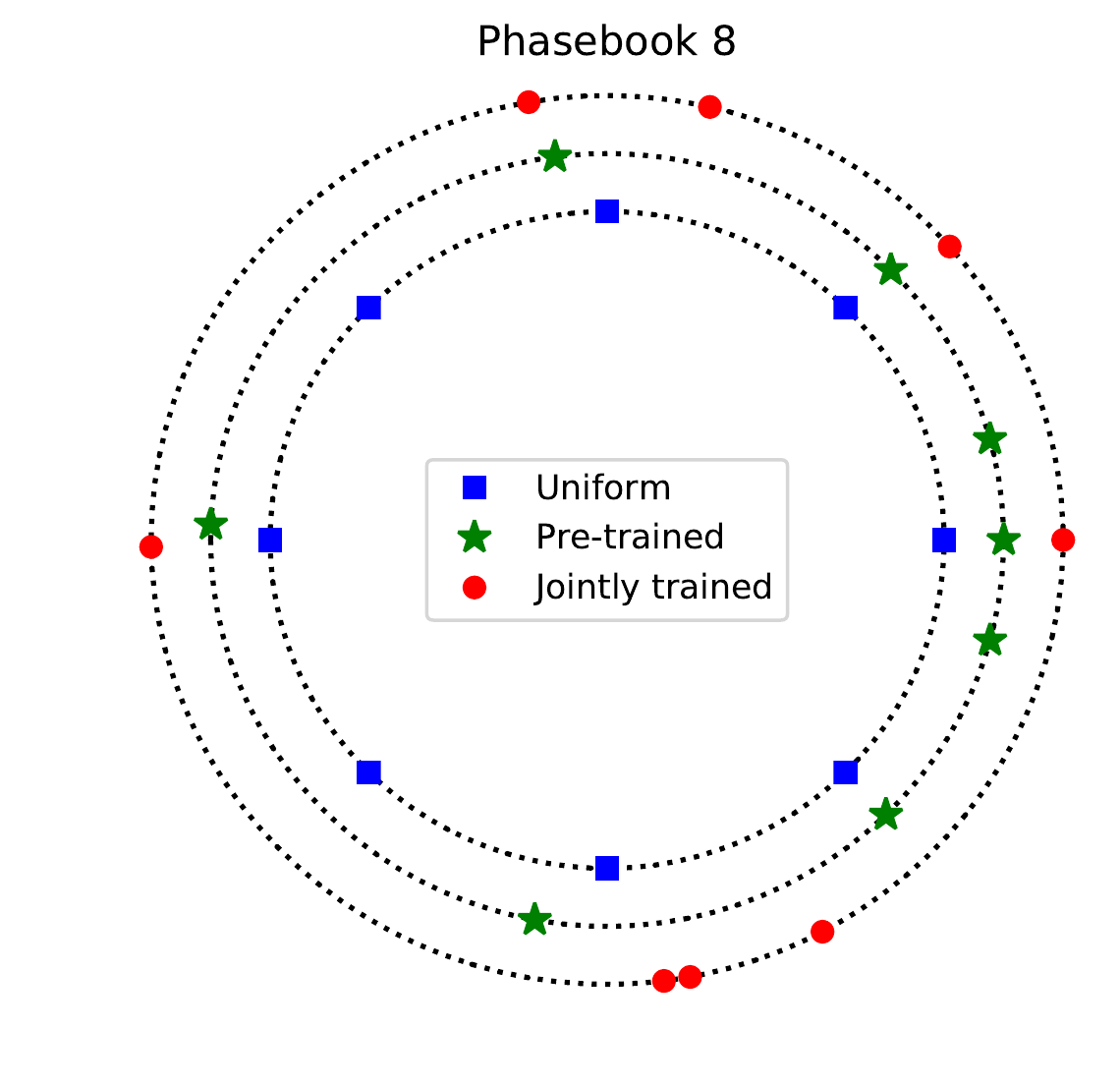}
		\includegraphics[width=.6\columnwidth]{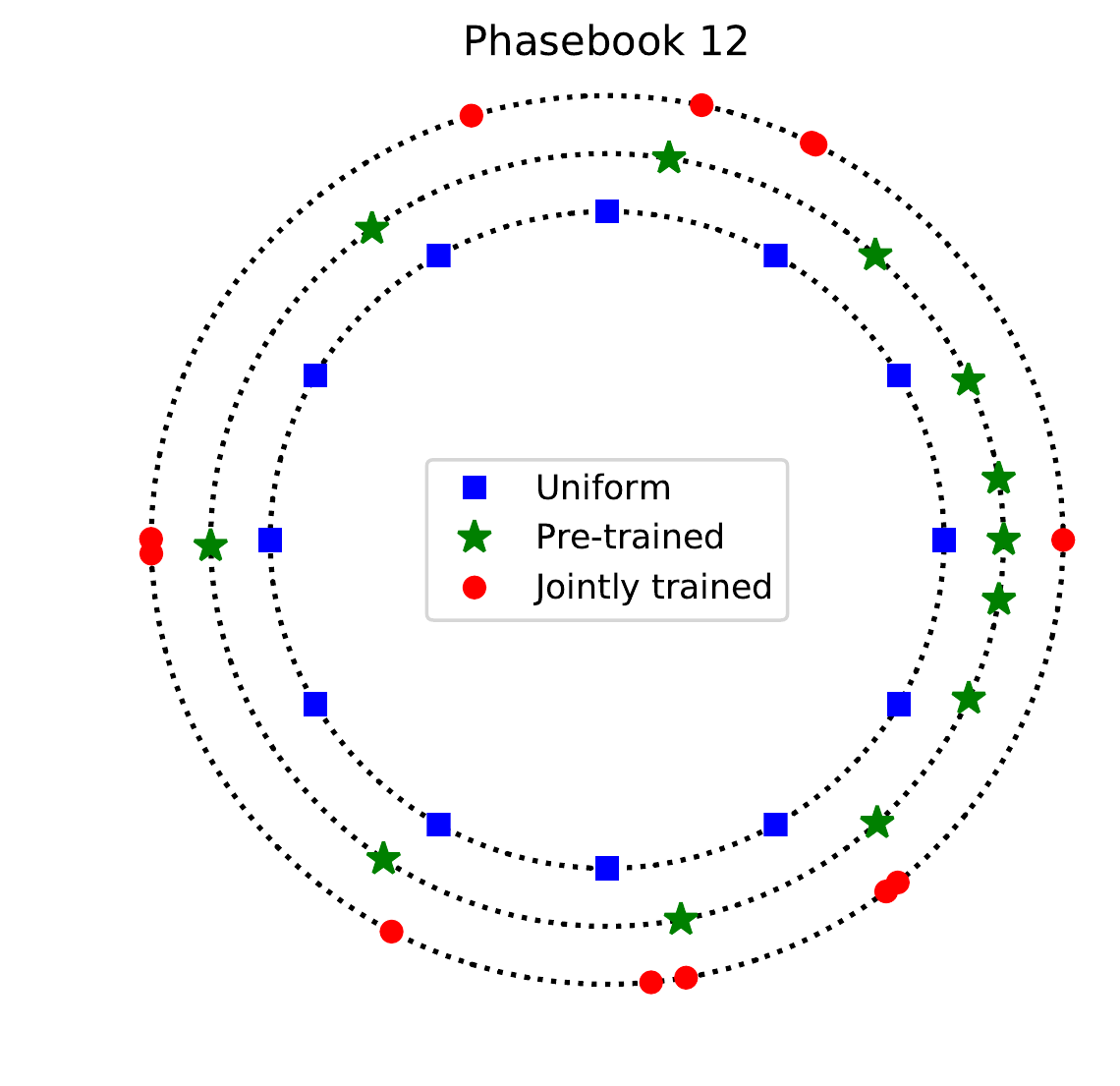}
	\caption{Uniform, pre-trained, and jointly trained phasebooks for $P\in\{4,812\}$; the pre-trained phasebooks are optimized assuming an oracle IAM estimate for magnitude, while the jointly trained phasebooks are optimized together with the rest of the network.}
	\label{fig:codebook_plot_size}
\end{figure*}

\begin{figure*}[htb]
	\centering
		\includegraphics[width=.49\columnwidth]{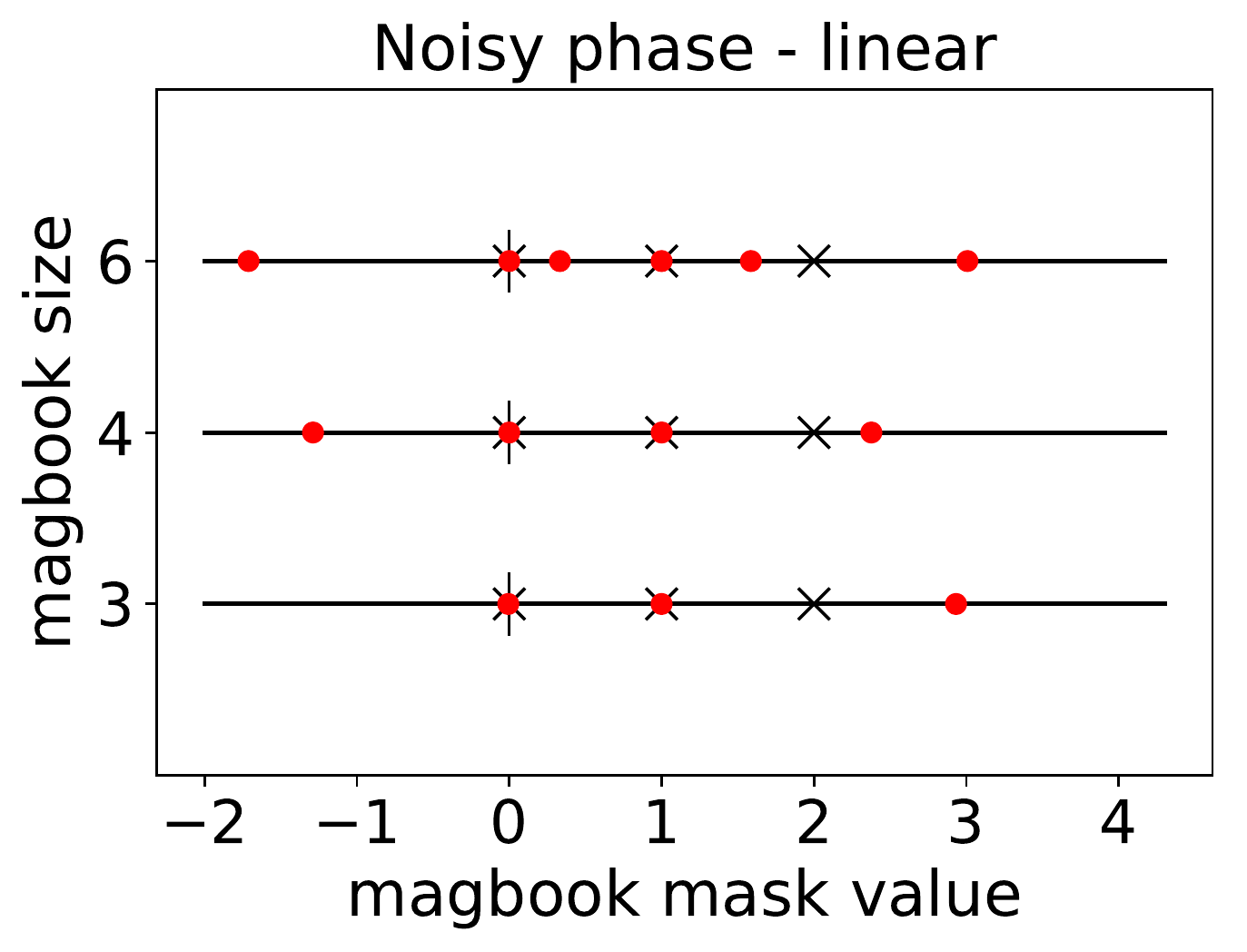}
		\includegraphics[width=.49\columnwidth]{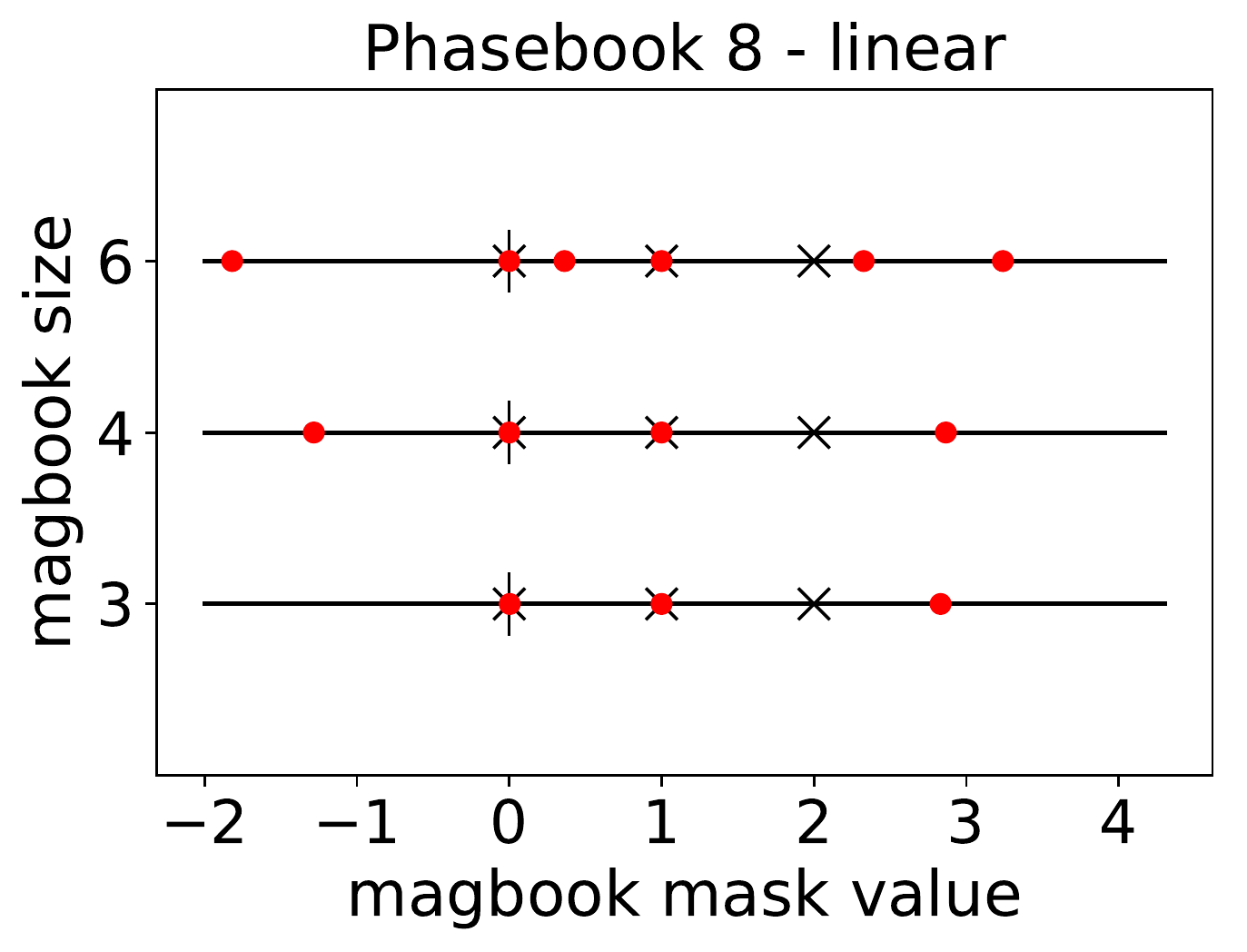}
		\includegraphics[width=.49\columnwidth]{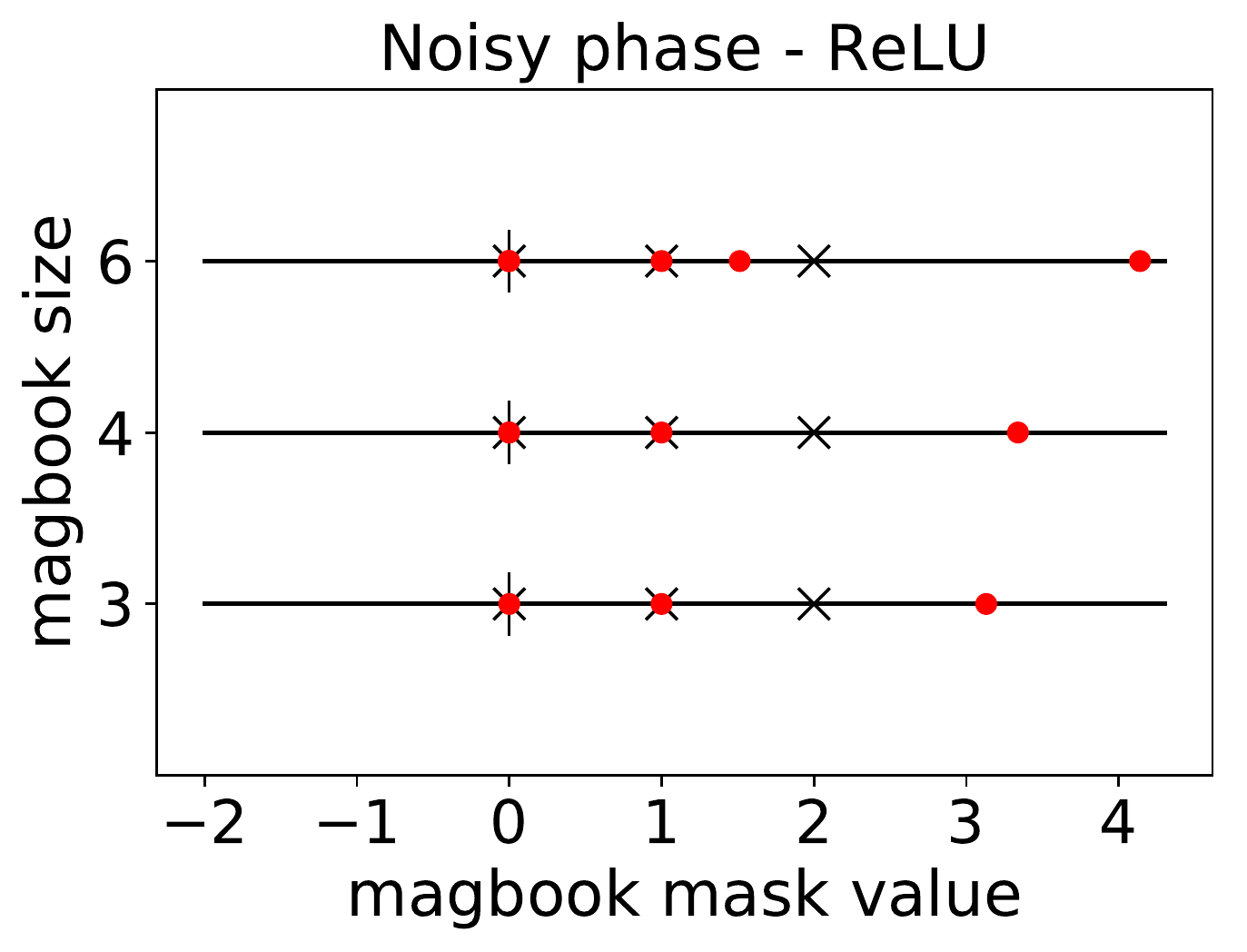}
		\includegraphics[width=.49\columnwidth]{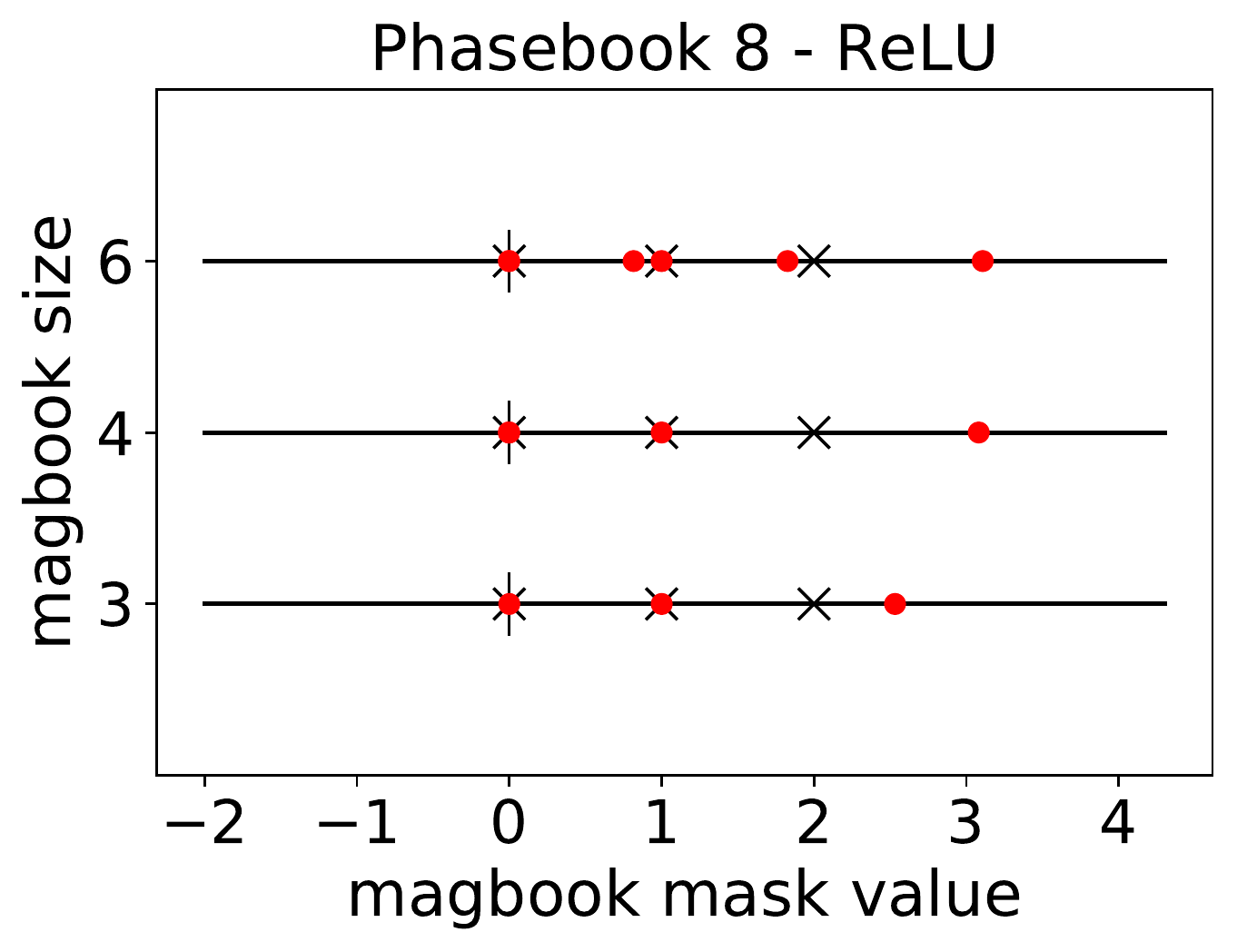}
	\caption{Jointly trained magbooks for different constraints on the magbook values (no constraint, as the output of a linear layer, or non-negative constraint, as the output of a ReLU layer) and different phase models (noisy phase or uniform phasebook layer with $P=8$ elements).  Red dots represent jointly trained magbook values, while crosses represent the fixed $\{0,1,2\}$ magbook, i.e., uniform magbook 3, as a reference.} 
	\label{fig:learn_magbooks}
\end{figure*}

\subsection{Influence of the phasebook size}

Figure~\ref{fig:bar_sdr_phasebook_type} shows 
SI-SDR improvements for various phasebook sizes, where phasebook values are either uniform, pre-trained offline assuming an oracle IAM magnitude, or jointly trained together with the rest of the network. In each case, both the magnitude mask and phase mask layers in the inference head are jointly fine-tuned using the WA loss function, after pre-training of a chimera++ network with WA loss on the MI head.  From Fig.~\ref{fig:bar_sdr_phasebook_type}, we see that all phasebooks improve on the noisy phase SI-SDR of 11.7 dB.  We also note that phasebooks of size 8 appear to perform best, and the uniform phasebooks perform comparably to those with learned values.  Note that, since we are interpolating over phasebook values, we can theoretically achieve the desired phase difference from any codebook, assuming it is dense enough, so the difference is mainly in the ease for the network to produce softmax outputs that are able to produce a correct estimate. We may see a different trend if we were to pick the argmax or to sample instead.

Figure~\ref{fig:codebook_plot_size} shows a comparison of the uniform phasebook with the pre-trained and jointly trained values for various codebook sizes.  We see that both the pre-trained and jointly trained values tend to place more weight between $-\pi/2$ and $+\pi/2$ as a majority of the learned values cluster in this range; this matches the empirical distribution shown in Fig.~\ref{fig:mask_statistics}.  We also note that the jointly trained phasebooks appear to be quite redundant, especially for $P=12$.

\subsection{Magbook}

We showed in \cite{Wang2018Interspeech09} that a convex softmax interpolation of fixed values $\{0,1,2\}$ for the magnitude mask leads to state-of-the-art performance when combined with an unfolded phase reconstruction algorithm. This corresponds to a uniform magbook 3 in our proposed framework. We here consider an extension of this case using the magbook formulation, where we further train end-to-end the values to be interpolated jointly with the softmax layer under a waveform approximation objective.

We consider two parameterizations for the magnitude: we let the parameters take any value in $\RR$ (``linear''), or we train them under a non-negative constraint, which we implemented using a ReLU non-linearity (``ReLU'').
We also consider two types of phase models: using the noisy phase as is, similarly to previous works, or using a phase mask obtained with a jointly trained phasebook layer with $P=8$ elements. All networks are first pre-trained from scratch as chimera networks then fine-tuned, each time using the WA objective on the MI head.

Figure~\ref{fig:learn_magbooks} shows examples of such learned magbooks. Interestingly, in the linear case, the network finds it best to use one or more negative magnitude elements: it is intuitive in the case of the noisy phase, where the network has an incentive to use its freedom to take negative values in order to fix the noisy phase in regions where a phase inversion is warranted; it is maybe slightly less intuitive when a phasebook layer is involved, as one may think that the phasebook layer should take care of phase inversions where they are needed instead of relying on negative magnitude mask values, but there is in fact no specific incentive in the objective function to favor a positive magnitude value $m$ associated with some phase $\theta$ versus the opposite magnitude value $-m$ with phase $\theta+\pi$, assuming both these phase values can be equally well generated by the phasebook layer.
In the ReLU case, the network can no longer use negative magnitudes, and tends to place multiple points close to $0$. To our surprise, it appears that magbooks obtained with the noisy phase featured slightly larger maximum values than those obtained with a phasebook layer, whereas we argued earlier that using the noisy phase should encourage the network to under-estimate the magnitude mask value. We plan to further investigate the behavior of the estimated masks in these cases by analyzing the estimated softmax probabilities and interpolated values.

Corresponding SI-SDR results are shown in Table~\ref{table:magbook}. We first observe that, when used together with the noisy phase, learning the magbook values appears slightly beneficial, especially when using the unconstrained (linear) magbooks, perhaps indicating that the network finds it useful to allocate some magbook values for phase inversion. However, when pairing the magnitude estimate with a better phase estimate obtained with a phasebook layer, learning the magbook values no longer brings improvements over the uniform magbook 3. The fact that there is little or no benefit in allowing the magbook layer to model a range other than $[0,2]$ is is in line with the oracle results shown in Fig.~\ref{fig:mask_comparison}, where truncation of the oracle IAM magnitude mask to $[0,2]$ brings significant benefits over the classical truncation to $[0,1]$, and truncation to a maximum value greater than $2$ brings little additional gain. The fact that the best results are obtained when interpolating between magbook values of $0$, $1$, and $R_\mathrm{max}$ (here equal to $2$) is in line with the true distribution of truncated magnitude mask values observed in Fig.~\ref{fig:mask_statistics}, with large peaks at $0$, $1$, and the truncation threshold $R_\mathrm{max}$.

\begin{table}[t]

    \footnotesize
	\centering
	\caption{SI-SDR improvements (dB) on the wsj0-2mix test set for various magbook sizes and nonlinearities.} %
	\label{table:magbook}
	\begin{tabular}{l|c|c}
		\hline\hline
		& \multicolumn{2}{c}{Phase estimate}\\
        Magnitude estimate & ~~~~~~Noisy~~~~~~ & Phasebook 8  \\ 
		\hline\hline
		Uniform magbook 3 & 11.7 & 12.4 \\
		\hline
		Jointly trained magbook 3 (linear)  & 11.9 & 12.2 \\
		\hline
		Jointly trained magbook 4 (linear)  & 12.1 & 12.2 \\
		\hline
		Jointly trained magbook 6 (linear)  & 12.1 & 12.4 \\
		\hline
		Jointly trained magbook 3 (ReLU) & 11.8 & 12.2 \\
		\hline
		Jointly trained magbook 4 (ReLU) & 11.8 & 12.3 \\
		\hline
		Jointly trained magbook 6 (ReLU) & 11.9 & 12.2 \\
		\hline\hline
	\end{tabular}%
\end{table}

\subsection{Combook}
\begin{figure}[t]
	\centering
		\includegraphics[width=.69\columnwidth]{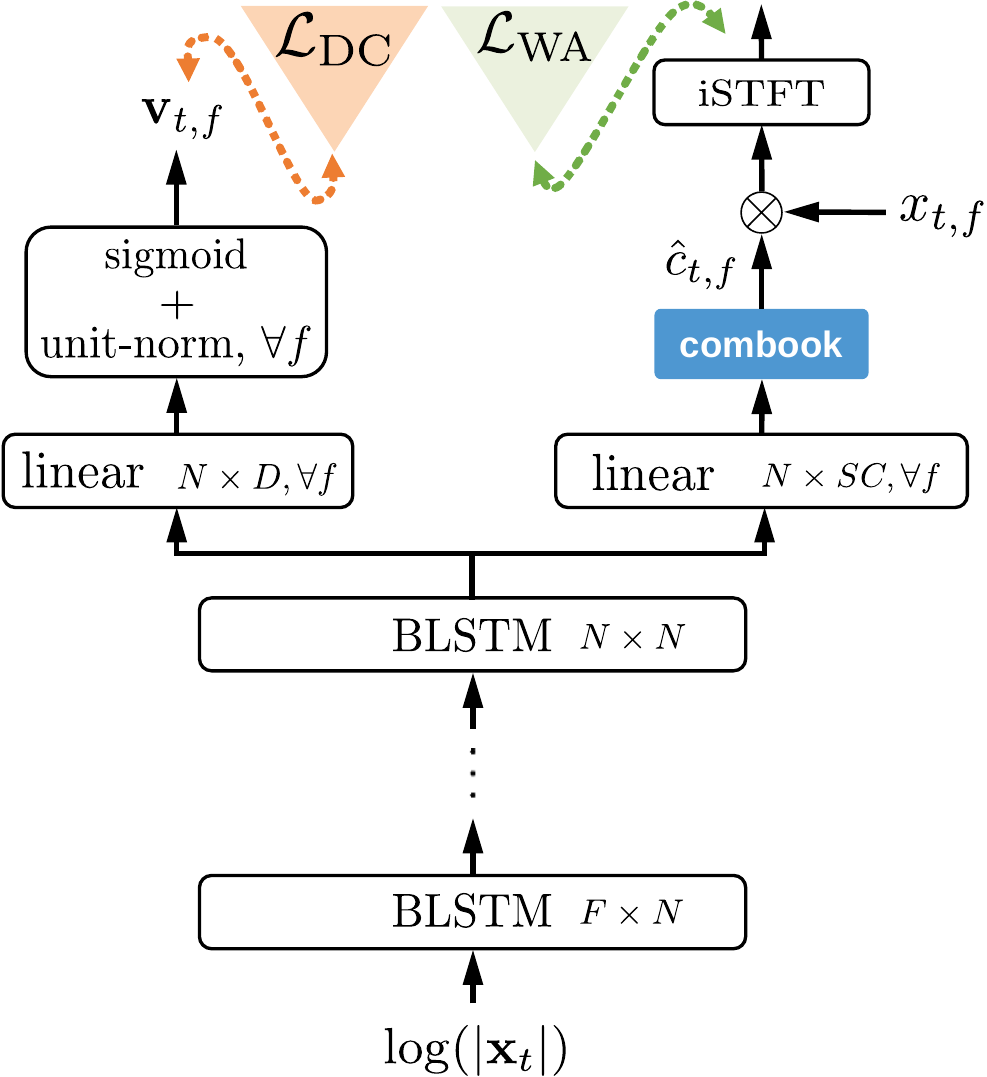}
	\caption{Chimera++ network with combook mask-inference head.}
	\label{fig:combook}
\end{figure}

\begin{figure*}[tb]
	\centering
		\includegraphics[width=.6\columnwidth]{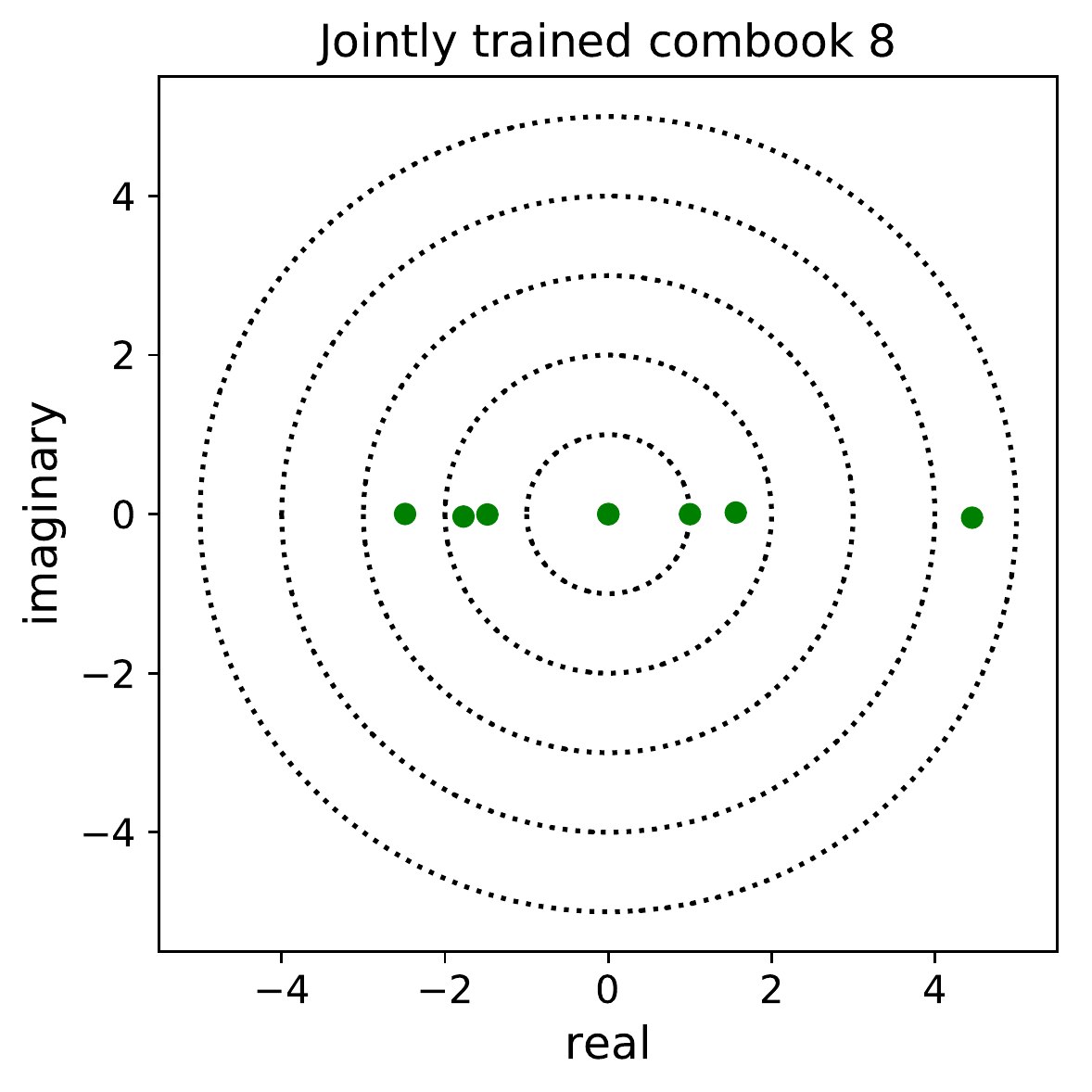}
		\includegraphics[width=.6\columnwidth]{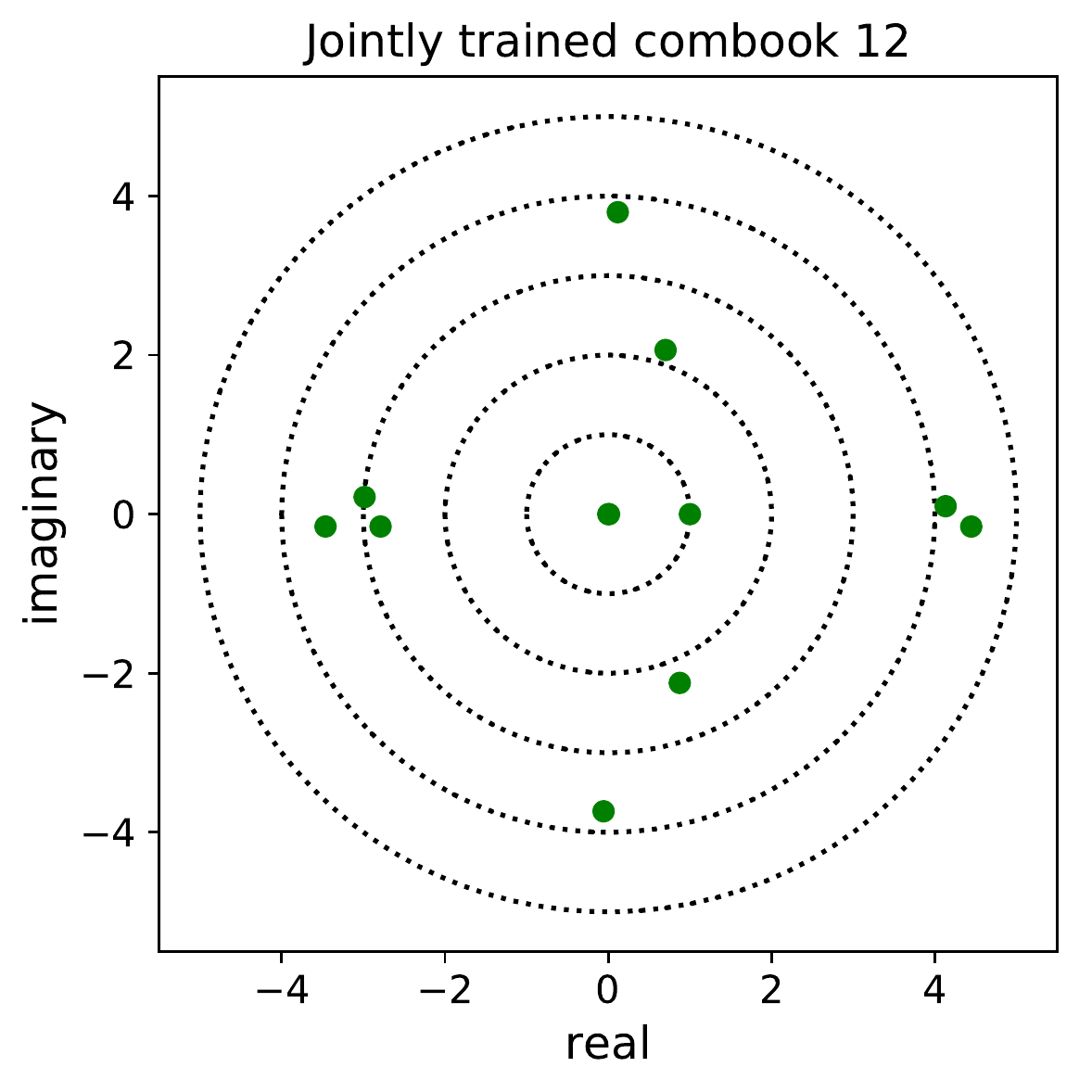}
		\includegraphics[width=.6\columnwidth]{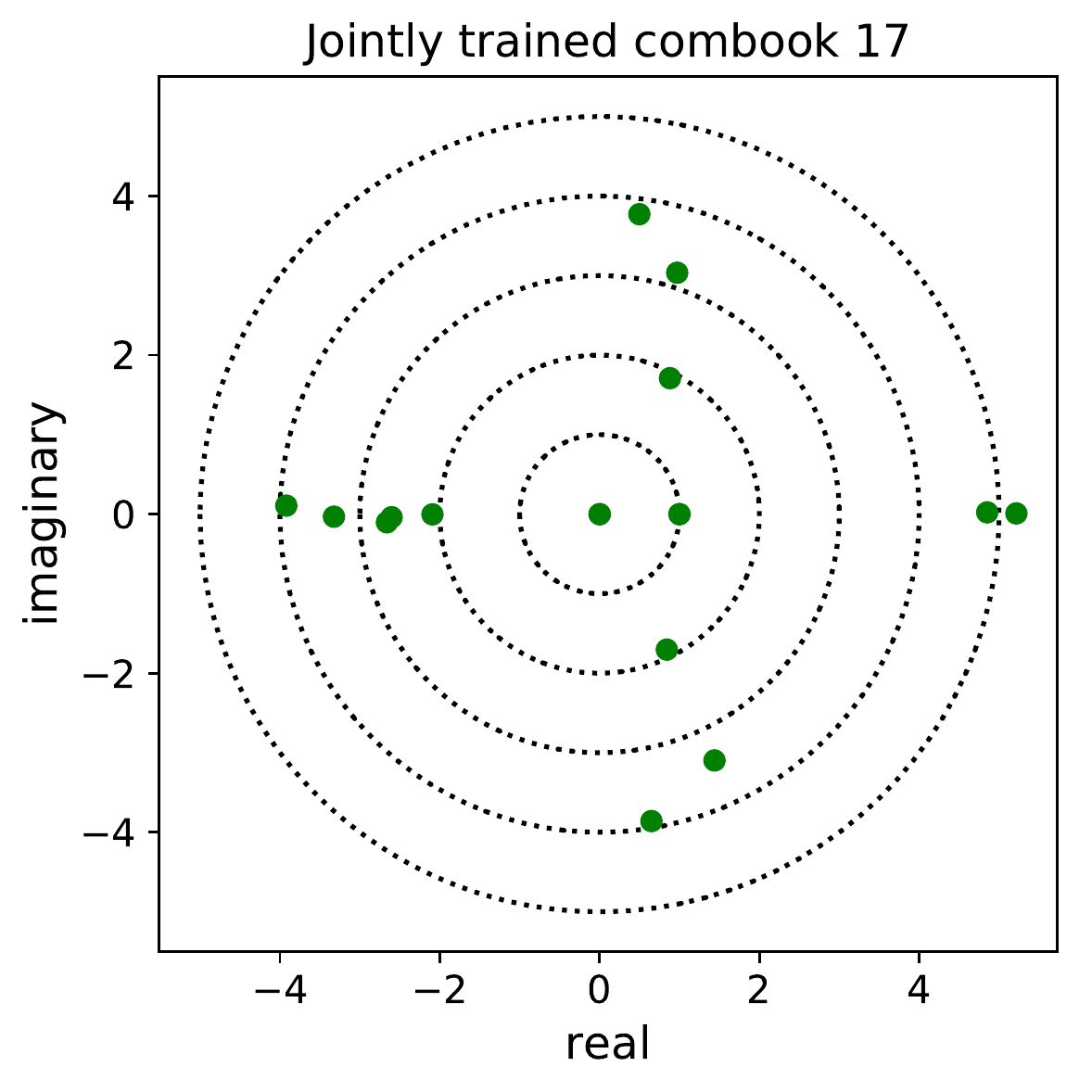}
	\caption{Jointly trained combooks for $C\in\{8,12,17\}$ for chimera++ training followed by mask-inference fine-tuning with WA objective.}
	\label{fig:learn_complex_codebook}
\end{figure*}

We have so far considered factorized representations of the complex mask as a product of a magnitude mask and a phase mask. We now consider a similar use of a discrete representation to model the complex mask, but directly using a codebook of complex values. We train Chimera++ networks where the magnitude mask estimation layer is replaced by a complex mask estimation layer consisting of a softmax layer used to interpolate values of a combook, as illustrated in Fig.~\ref{fig:combook}. The networks are trained from scratch with both deep clustering and WA objectives, then fine-tuned with WA objective only.

Examples of learned combooks are shown in Fig.~\ref{fig:learn_complex_codebook} for $C\in\{8,12,17\}$.  We note that the combook size $C$ should not be directly compared to the phasebook size $P$ and magbook size $M$ of the previous sections, since the phasebook and magbook combine to lead to complex values: in the argmax scheme, setting aside one magbook (and combook) value which will most likely be at 0, we have $P$ phasebook values for each of the remaining $M-1$ magbook values.  A given combination of magbook and phasebook values can thus be considered similar to a set of combook values of size $C=1+P(M-1)$, e.g., $(M,P)=(3,8)$ is akin to $C=17$.  Interestingly, for small sizes such as $C=8$, the combook layer does not take advantage of non-real values, focusing first on covering negative values (for phase inversion), $0$, and positive values. This is similar to what we observe with some of the linear magbooks in Fig.~\ref{fig:learn_magbooks} that learn to allocate magnitude values for phase inversion.  Only with $C=12$ in Fig.~\ref{fig:learn_complex_codebook} do we start seeing non-real values. We note however that the network does not appear to be very efficient in its usage of the available values, learning seemingly redundant values, such as %
the cluster of points near $-3+0j$ in the middle and far right plots of Fig.~\ref{fig:learn_complex_codebook}.

Table~\ref{table:combook} compares SI-SDR results for  combooks of various sizes, in addition to the best performing magbook and phasebook configurations.  It appears that, in the current setup, the ability of the combook layer to estimate a complex mask via a single network layer works slightly better than trying to estimate magnitude and phase via separate layers.  Table~\ref{table:combook} also shows that performance does not improve for combook sizes greater than $C=12$, which we use going forward.

\begin{table}[t]

    \footnotesize
	\centering
	\caption{SI-SDR improvements (dB) on the wsj0-2mix test set for various combook sizes.  Best magbook and phasebook results are also shown.} %
	\label{table:combook}
	\begin{tabular}{l|c}
		\hline\hline
        Codebook & SI-SDR (dB)  \\ 
		\hline\hline
		Jointly trained combook 4 & 12.1  \\
		\hline
		Jointly trained combook 8  & 12.1  \\
		\hline
		Jointly trained combook 12  & 12.6  \\
		\hline
		Jointly trained combook 17  & 12.5  \\
		\hline
		Jointly trained combook 24  & 12.6  \\
		\hline
		Jointly trained magbook 4 w/ noisy phase & 12.1  \\
		\hline
		Uniform magbook 3 w/ uniform phasebook 8  & 12.4  \\
		\hline\hline
	\end{tabular}%
\end{table}

\begin{figure}[t]
	\centering
		\includegraphics[width=.9\columnwidth]{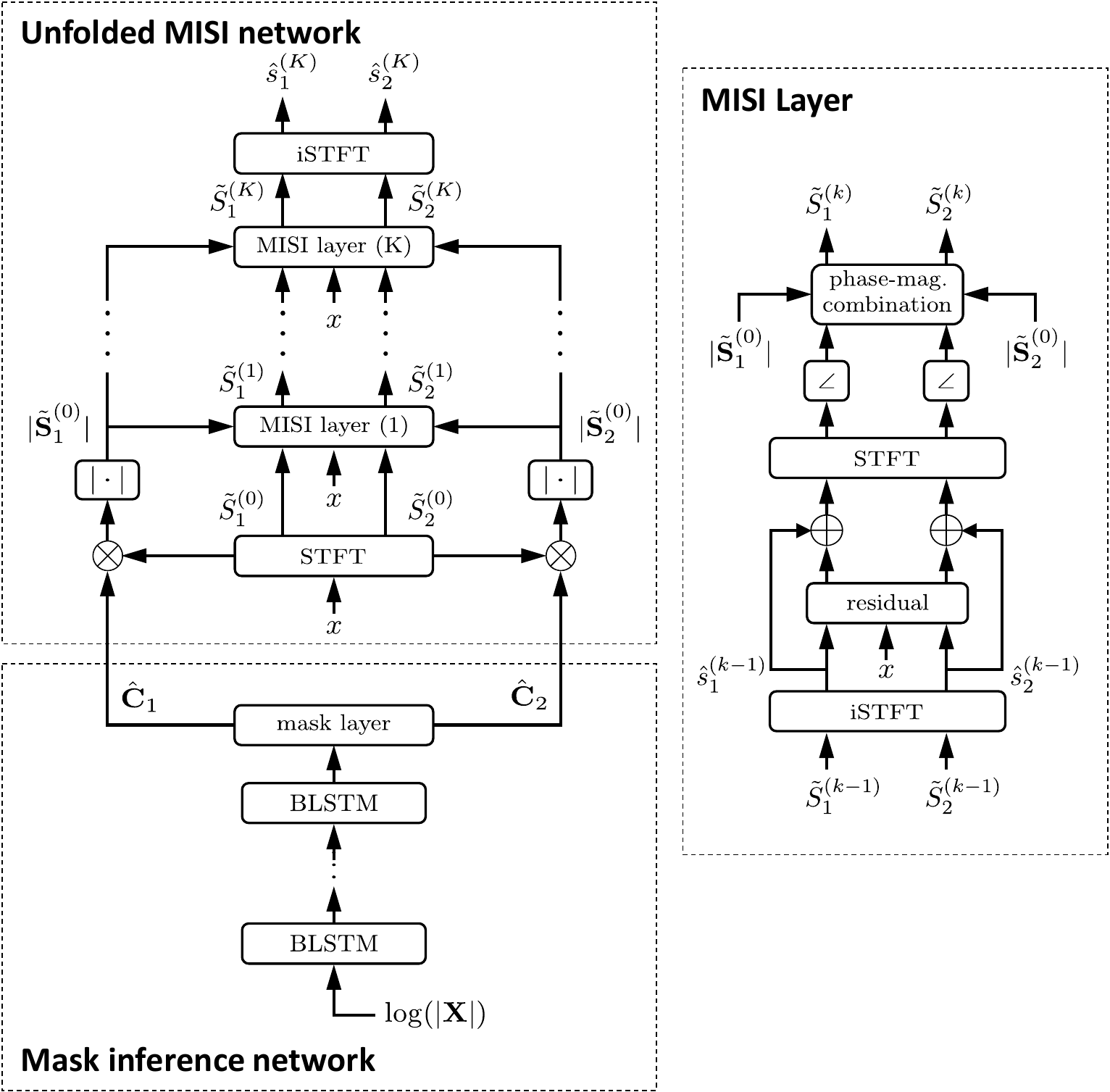}
	\caption{Mask inference part of a Chimera++ network with unfolded MISI reconstruction.}
	\label{fig:misi-network}
\end{figure}

\subsection{Training through unfolded MISI}

Following \cite{Wang2018Interspeech09}, we now consider adding an unfolded MISI network with $K$ iterations at the output of the MI head, as illustrated in Fig.~\ref{fig:misi-network}, and training the full network using the WA-MISI-K loss function. In the figure, the masks $\hat{\bf C}_i$ shall be considered as complex when phasebook or combook layers are involved, and as real otherwise.

\begin{figure}[t]%
	\centering
		\includegraphics[width=.99\columnwidth]{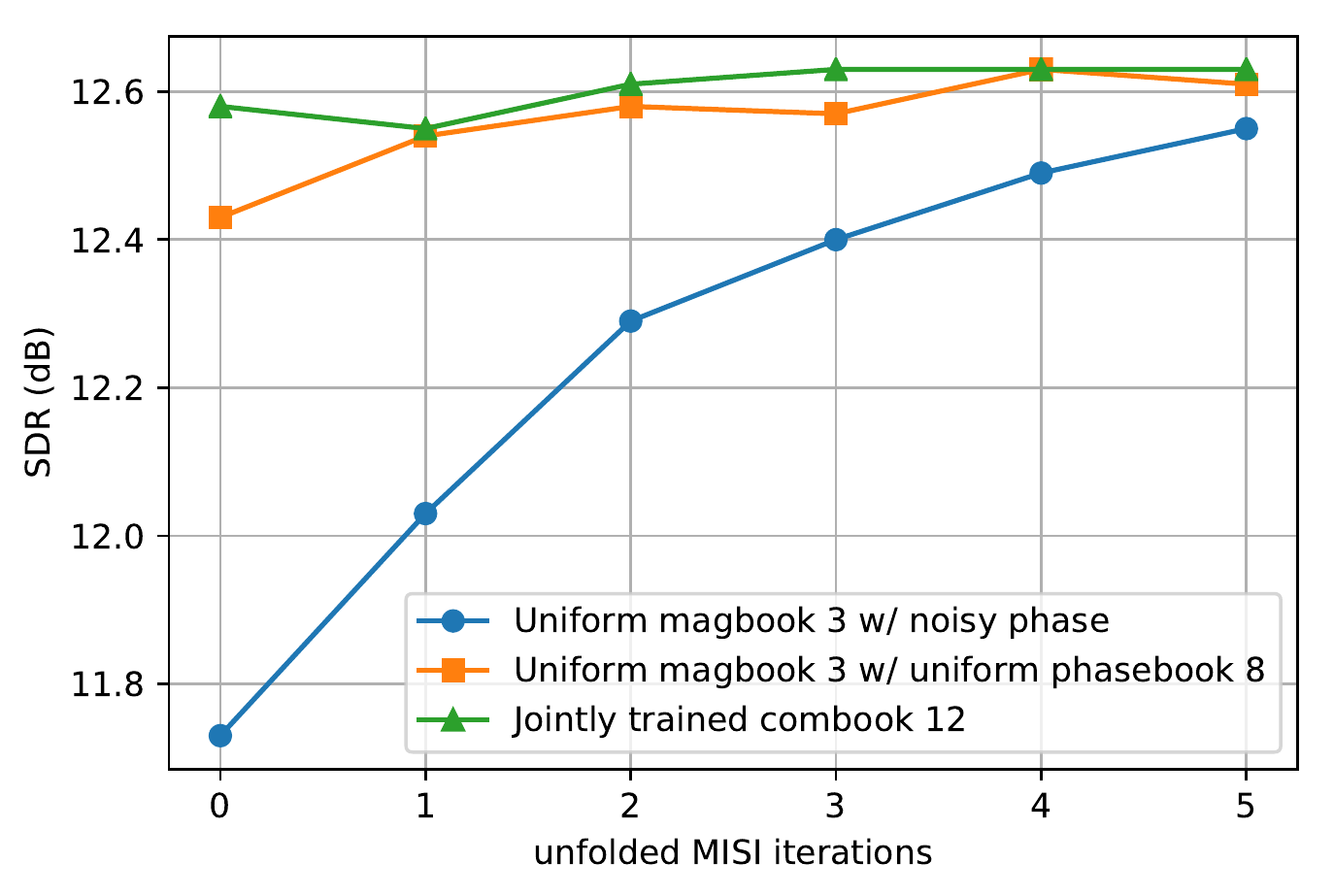}%
	\caption{SI-SDR improvement (dB) for a given number of unfolded MISI iterations from a complex time-frequency domain speech estimate obtained by: (1) combining an estimated magnitude mask from a uniform magbook 3 layer  with the noisy phase; (2) combining an estimated magnitude mask from a uniform magbook 3 layer with the phase mask obtained with a uniform phasebook 8 layer; and (3) using a complex mask obtained with a Jointly trained combook 12 layer trained jointly with the rest of the network.}
	\label{fig:MISI_sdr_codebook_type}
\end{figure}

Results are shown in Fig.~\ref{fig:MISI_sdr_codebook_type} for various numbers of unfolded MISI iterations, and three different types of networks: the original chimera++ network using the noisy phase with a uniform magbook 3 layer with fixed elements $\{0,1,2\}$, as a (state-of-the-art) baseline; a chimera++ network with the same architecture and an additional phasebook layer with $P=8$ uniformly distributed elements; a chimera++ network with a combook layer as MI head whose $C=12$ elements are learned end-to-end together with the rest of the network parameters. We observe that the combook network improves significantly over the noisy phase baseline and obtains the best performance among all methods for direct iSTFT reconstruction (i.e., $0$ MISI iterations), but its performance does not further improve. The phasebook network also improves significantly over the baseline, and converges to an SI-SDR value similar to that of the combook in $K=2$ iterations. Both combook and phasebook enable a better phase estimate which can match state-of-the-art performance without the need for unfolded phase reconstruction required when using the noisy phase.

Table~\ref{tab:compare} shows a comparison of the best proposed systems with three recently proposed approaches: the original Chimera++ network using noisy phase and MISI phase reconstruction as a post-processing only \cite{Wang2018ICASSP04Alternative}; a Chimera++ network trained through unfolded MISI phase reconstruction \cite{Wang2018Interspeech09}, which is equivalent in our framework to a uniform magbook 3 with noisy phase as the initial phase; and a Chimera++ network with unfolded phase reconstruction in which the STFT and iSTFT transforms are replaced by separate (or ``untied'') transforms at each layer, learned together with the rest of the network \cite{Wichern2018IWAENC09}. The Jointly trained combook 12 system obtains the best performance when no MISI iteration is performed, at 12.6 dB, beating the previous state-of-the-art 12.2 dB which involves further learning a transform replacing the final iSTFT \cite{Wichern2018IWAENC09}. If we allow ourselves 5 MISI iterations, all proposed systems reach 12.6 dB, but they are slightly outperformed by the system which learns replacements for the STFT/iSTFT transforms, with 12.8 dB. We shall leave it to future work to combine such transform learning with our proposed systems.

{\setlength{\tabcolsep}{5pt}
\begin{table}[t]%

    \footnotesize
	\centering
	\caption{SI-SDR (dB) comparison with other recent systems on the wsj0-2mix test set.}
	\label{tab:compare}
	\begin{tabular}{l|c|c}
		\hline\hline
		& MISI & \!SI-SDR\! \\
        Approach & \!Iterations\! & [dB]   \\ 
		\hline\hline
		Chimera++ \cite{Wang2018ICASSP04Alternative} & 0 & 11.2  \\
		& 5 & 11.5  \\ \hline
		Uniform magbook 3 w/ noisy phase \cite{Wang2018Interspeech09} & 0 & 11.8  \\
		& 5 & 12.6  \\ \hline
		Unfolded MISI with learned untied transforms \cite{Wichern2018IWAENC09}& 0 & 12.2  \\
		 & 5 & {\bf 12.8}  \\ \hline
		Uniform magbook 3 w/ uniform phasebook 8 & 0 & 12.4  \\
		& 5 & 12.6  \\ \hline
		Jointly trained combook 12 & 0 & {\bf 12.6}  \\
		& 5 & 12.6  \\ 
		\hline\hline
	\end{tabular}%
\end{table}
}

\section{Conclusion and future works}

According to the above experiments, both a combook layer and a combination of magbook and phasebook layers can significantly improve the performance of single-channel multi-speaker speech separation, especially reducing the need for further phase reconstruction. We have here focused mostly on end-to-end training using the waveform approximation objective, because it has led to the best results both here and in recent work \cite{Wang2018Interspeech09}: the most convenient way to use this objective for magbook, phasebook, and combook layers was to rely on the interpolation scheme, where our losses are computed on the expected outputs over the codebooks. We could also investigate training through the argmax scheme by considering expected loss functions that compute the expectation of the loss over each possible value in the codebook. 
This would in particular allow us to use the discrete nature of the representation to introduce conditional probability relationships between T-F bins. However, as mentioned in Section~\ref{sec:expected_loss}, such expectations are typically intractable, necessitating methods such as the Gumbel-Softmax \cite{jang2016categorical,maddison2016concrete} or the policy gradient technique \cite{williams1992simple}.
\if 0
We also plan to investigate the performance of the expected losses introduced in Section~\ref{sec:expected_loss}, based on marginalizing the CSA loss or sampling the WA loss; we did not explore these here as they are more involved than the vanilla WA loss. 
\else
\fi
Finally, while we here considered estimating the difference between the noisy and clean phase, we can consider also estimating the clean phase directly, and train the network to merge the two estimates based on the context.
\\

\bibliographystyle{IEEEtran_nourl}
\bibliography{phasebook}

\begin{IEEEbiography}[{\includegraphics[width=1in,height=1.25in,clip,keepaspectratio]{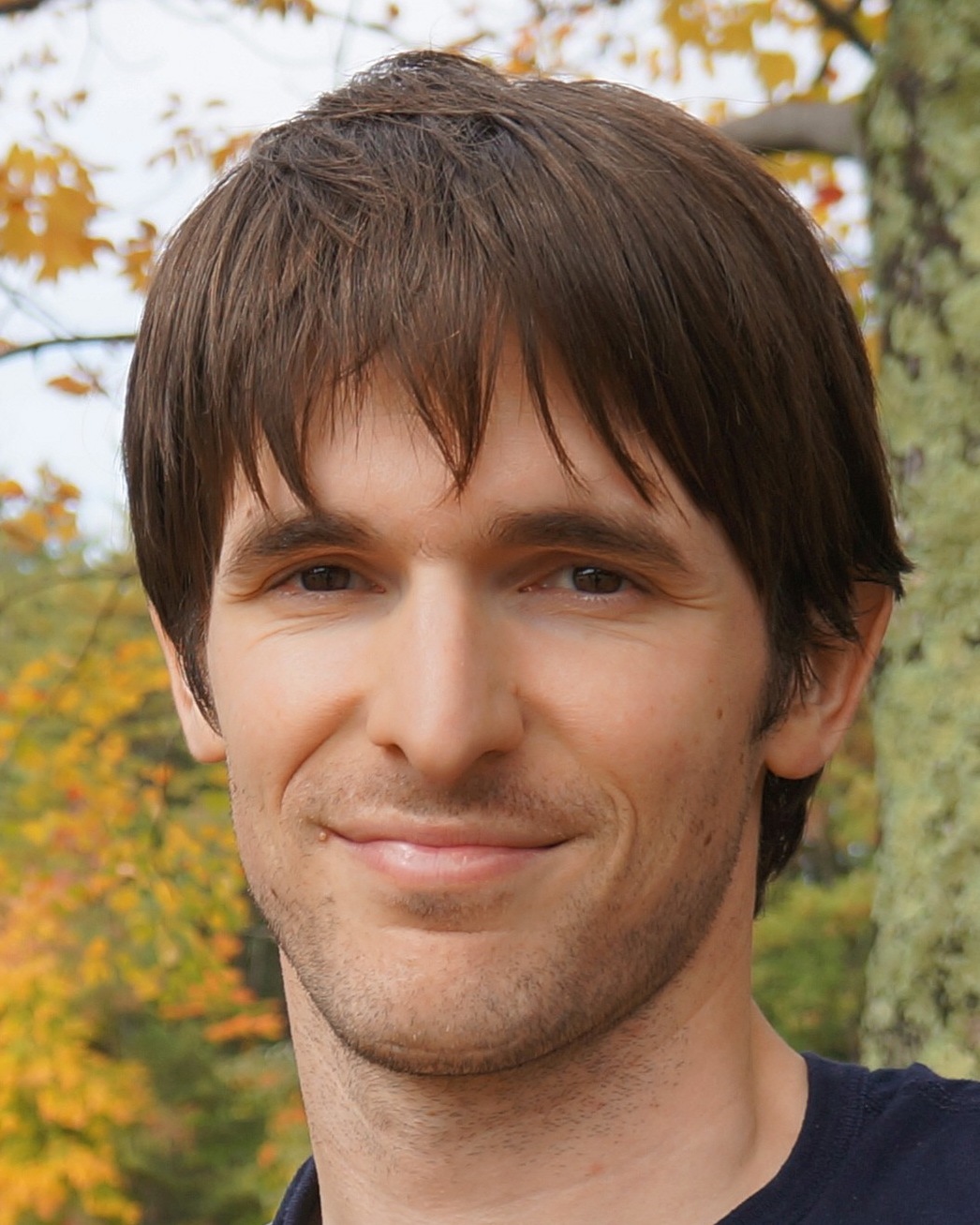}}]{Jonathan Le Roux} is a Senior Principal Research Scientist and the Speech and Audio Team Leader at Mitsubishi Electric Research Laboratories (MERL) in Cambridge, Massachusetts. He completed his B.Sc.\ and M.Sc.\ degrees in Mathematics at the Ecole Normale Sup{\'e}rieure (Paris, France), his Ph.D.\ degree at the University of Tokyo (Japan) and the Universit{\'e} Pierre et Marie Curie (Paris, France), and worked as a postdoctoral researcher at NTT's Communication Science Laboratories from 2009 to 2011. His research interests are in signal processing and machine learning applied to speech and audio. He has contributed to more than 80 peer-reviewed papers and 20 patents in these fields. He is a founder and chair of the Speech and Audio in the Northeast (SANE) series of workshops, a Senior Member of the IEEE and a member of the IEEE Audio and Acoustic Signal Processing Technical Committee (AASP).
\end{IEEEbiography}

\begin{IEEEbiography}[{\includegraphics[width=1in,height=1.25in,clip,keepaspectratio]{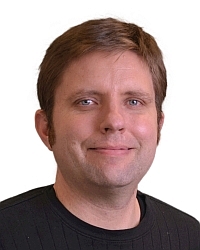}}]{Gordon Wichern} is a Principal Research Scientist at Mitsubishi Electric Research Laboratories (MERL) in Cambridge, Massachusetts. He received his B.Sc.\ and M.Sc.\ degrees from Colorado State University in electrical engineering and his Ph.D.\ from Arizona State University in electrical engineering with a concentration in arts, media and engineering, where he was supported by a National Science Foundation (NSF) Integrative Graduate Education and Research Traineeship (IGERT) for his work on environmental sound recognition.  He was previously a member of the research team at iZotope, inc.\ where he focused on applying novel signal processing and machine learning techniques to music and post production software, and a member of the Technical Staff at MIT Lincoln Laboratory where he worked on radar signal processing.  His research interests include audio, music, and speech signal processing, machine learning, and psychoacoustics.
\end{IEEEbiography}

\begin{IEEEbiography}[{\includegraphics[width=1in,height=1.25in,clip,keepaspectratio]{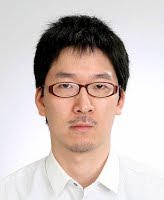}}]{Shinji Watanabe} is an Associate Research Professor at Johns Hopkins University, Baltimore, MD. He received his B.S., M.S., and PhD Degrees in 1999, 2001, and 2006, respectively, all from Waseda University, Tokyo, Japan. He was a research scientist at NTT Communication Science Laboratories, Kyoto, Japan, from 2001 to 2011, a visiting scholar in Georgia institute of technology, Atlanta, GA in 2009, and a Senior Principal Research Scientist at Mitsubishi Electric Research Laboratories (MERL), Cambridge, MA from 2012 to 2017. His research interests include machine learning and speech and spoken language processing. He has been published more than 150 papers in journals and conferences, and received several awards including the best paper award from the IEICE in 2003. He served as an Associate Editor of the IEEE Transactions on Audio Speech and Language Processing, and is a member of several technical committees including the IEEE Signal Processing Society Speech and Language Technical Committee (SLTC) and Machine Learning for Signal Processing Technical Committee (MLSP).
\end{IEEEbiography}

\begin{IEEEbiography}[{\includegraphics[width=1in,height=1.25in,clip,keepaspectratio]{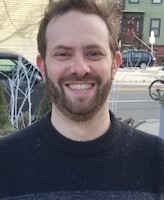}}]{Andy Sarroff} is a Research Engineer at iZotope, Inc. in Cambridge, MA, USA. He contributed to this publication while performing as a Research Intern at Mitsubishi Electric Research Laboratories (MERL) in Cambridge, MA, USA. Andy has received a BA in Music from Wesleyan University (2000); an MM in Music Technology from New York University (2009); and a PhD in Computer Science from Dartmouth College (2018). Andy has been a visiting researcher at Columbia University's Laboratory for the Recognition and Organization of Speech and Audio (LabROSA; 2015) and served on the board of directors of the International Society for Music Information Retrieval (ISMIR; 2015-2017). Andy's research interests include machine learning, machine listening and perception, and signal processing.
\end{IEEEbiography}

\begin{IEEEbiography}[{\includegraphics[width=1in,height=1.25in,clip,keepaspectratio]{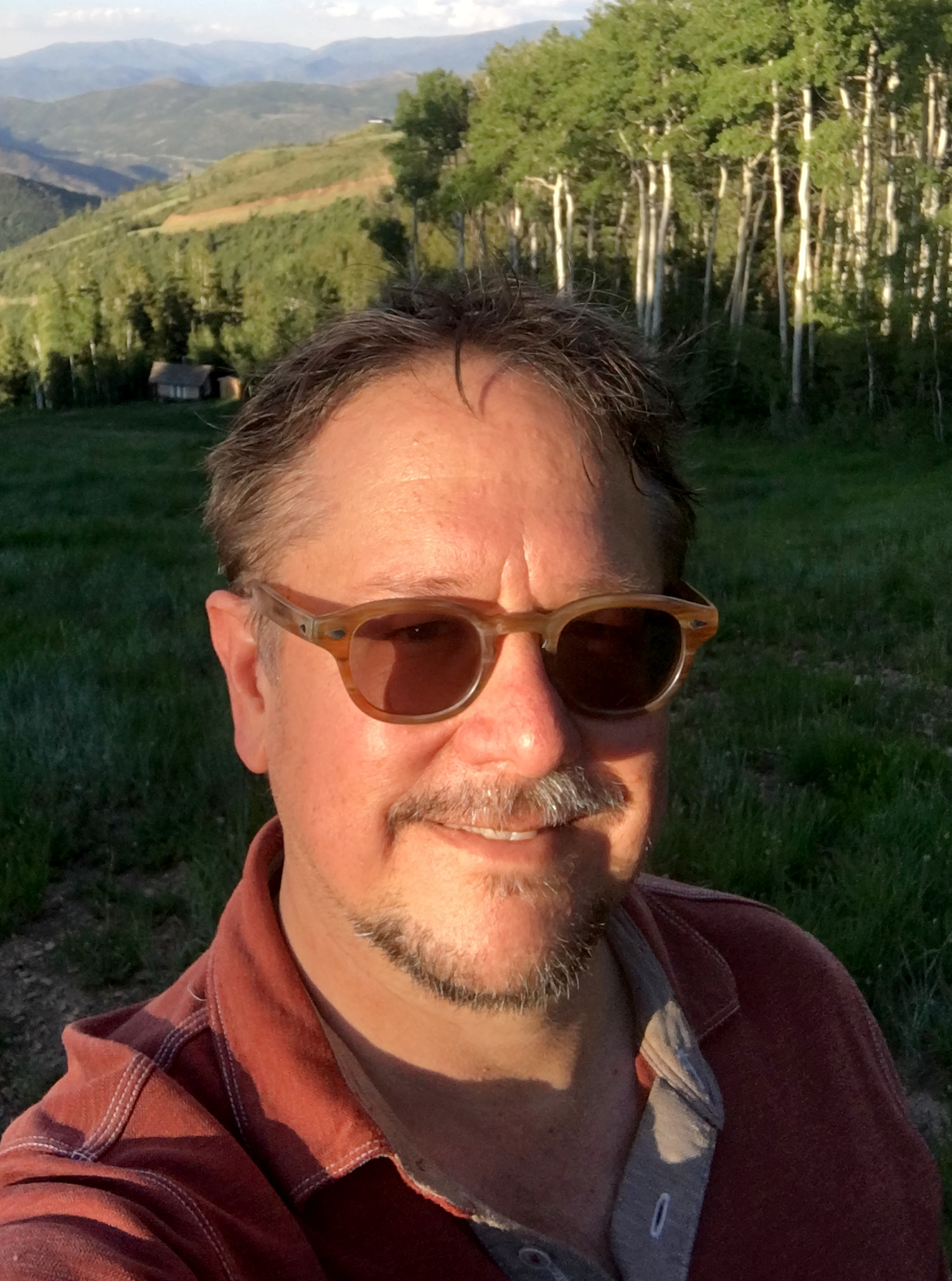}}]{John R. Hershey} is a researcher in Google AI Perception in Cambridge, Massachusetts, where he leads a research team in the area of speech and audio machine perception. Prior to Google, he spent seven years leading the speech and audio research team at MERL (Mitsubishi Electric Research Labs), and five years at IBM's T. J. Watson Research Center in New York, where he led a team of researchers in noise-robust speech recognition. He also spent a year as a visiting researcher in the speech group at Microsoft Research in 2004, after obtaining his Ph.D.\ from UCSD. Over the years, he has contributed to more than 100 publications and over 30 patents in the areas of machine perception, speech and audio processing, audio-visual machine perception, speech recognition, and natural language understanding.
\end{IEEEbiography}

\end{document}